\documentclass[a4paper,onecolumn,11pt,accepted=2023-03-21]{quantumarticle}
\pdfoutput=1
\usepackage{graphicx,tabularx}
\usepackage{amsmath,amssymb,amsfonts,amsthm,mathtools}
\usepackage{CJK,xparse,physics,color,bm}
\usepackage{ulem}

\usepackage[utf8]{inputenc}
\usepackage[colorlinks=true,breaklinks=true,allcolors=blue]{hyperref}

\usepackage[numbers,sort&compress]{natbib}
\usepackage[english]{babel}
\usepackage[T1]{fontenc}
\usepackage{tikz}
\usepackage{lipsum}

\begin{document}
\title{General properties of fidelity in non-Hermitian quantum systems with PT symmetry}
\author{Yi-Ting Tu}
\affiliation{Department of Physics, University of Maryland, College Park, MD, USA}
\author{Iksu Jang}
\affiliation{Department of Physics, National Tsing Hua University, Hsinchu 300044, Taiwan}
\author{Po-Yao Chang}
\affiliation{Department of Physics, National Tsing Hua University, Hsinchu 300044, Taiwan}
\author{Yu-Chin Tzeng}
\orcid{0000-0002-0380-1431}
\email{yctzeng@phys.ncts.ntu.edu.tw}
\affiliation{Physics Division, National Center for Theoretical Sciences, Taipei 106319, Taiwan}
\affiliation{Center for Theoretical and Computational Physics, National Yang Ming Chiao Tung University, Hsinchu 300093, Taiwan}
\begin{abstract}
  The fidelity susceptibility is a tool for studying quantum phase transitions in the Hermitian condensed matter systems. Recently, it has been generalized with the biorthogonal basis for the non-Hermitian quantum systems. From the general perturbation description with the constraint of parity-time (PT) symmetry, we show that the fidelity $\mathcal{F}$ is always real for the PT-unbroken states. For the PT-broken states, the real part of the fidelity susceptibility $\mathrm{Re}[\mathcal{X}_F]$ is corresponding to considering both the PT partner states, and the negative infinity is explored by the perturbation theory when the parameter approaches the exceptional point (EP). Moreover, at the second-order EP, we prove that the real part of the fidelity between PT-unbroken and PT-broken states is $\mathrm{Re}\mathcal{F}=\frac{1}{2}$. Based on these general properties, we study the two-legged non-Hermitian Su-Schrieffer-Heeger (SSH) model and the non-Hermitian XXZ spin chain. We find that for both interacting and non-interacting systems, the real part of fidelity susceptibility density goes to negative infinity when the parameter approaches the EP, and verifies it is a second-order EP by $\mathrm{Re}\mathcal{F}=\frac{1}{2}$.
\end{abstract}
\maketitle
\tableofcontents
\section{Introduction}\label{sec:motivation}
Quantum phase transition is one of the main interests in many-body systems. Instead of thermal fluctuation, the quantum phase transition which takes place at zero temperature is caused by quantum fluctuation. By tuning the external parameter strength $\lambda$, the ground-state $|\psi_0(\lambda)\rangle$ of a  Hermitian system is expected to have a drastic change and to undergo the quantum phase transition at the quantum critical point $\lambda_c$. Therefore, a simple hypothesis for detecting the quantum critical point is that there exists a drastic drop in the fidelity around the critical point. The fidelity $\mathcal{F}_{\rm h}(\lambda)=\abs{\langle\psi_0(\lambda)|\psi_0(\lambda+\epsilon)\rangle}^2$ is defined by the inner-product of the two ground-states with nearby parameters $\lambda$ and $\lambda+\epsilon$. This drastic drop can be characterized by a divergent quantity called the fidelity susceptibility $\mathcal{X}_{F_{\rm h}}$. Because the first-order term in the expansion of the fidelity vanishes, $\mathcal{F}_{\rm h}=1-\mathcal{X}_{F_{\rm h}}\epsilon^2+O(\epsilon^3)$, the fidelity susceptibility is the second-order coefficient and can be approximated by $\mathcal{X}_{F_{\rm h}}\approx(1-\mathcal{F}_{\rm h})/\epsilon^2$ for computation.
The fidelity susceptibility has been used to probe quantum phase transitions for more than a decade~\cite{You2007,fidelity_review}. 
Recent advances in fidelity, on the one hand, are studying higher-order phase transitions in the Hermitian systems,~\cite{MFYang2007,Fj_restad_2008, Tzeng2008a,Tzeng2008b, Ren_2015, fsus_BKT_2010, fsus_BKT_2011, fsus_BKT_2015, zhang2021fidelity}; and on the other hand, are the extension to the non-Hermitian systems.

There is a wide variety of classical and quantum systems described by non-Hermitian matrices/operators from different kinds of the physical origins of non-Hermiticity~\cite{review-Ashida2020,Ganainy2018}. 
For example, the Lindblad master equation~\cite{Lindblad1976} describing the dissipation in open quantum systems reduces into the Schr\"odinger-like equation with a non-Hermitian effective Hamiltonian. Because of $H^\dagger{\neq}H$, the time evolution of wavefunctions is driven by both $H$ and $H^\dagger$, simultaneously. We have $\frac{\partial}{\partial t}|\varphi^R(t)\rangle{=}{-}iH|\varphi^R(t)\rangle$, and $\frac{\partial}{\partial t}|\varphi^L(t)\rangle{=}{-}iH^\dagger|\varphi^L(t)\rangle$. Naturally, as the standard linear algebra, the eigenvectors of the non-Hermitian Hamiltonian are generalized into biorthogonal left and right eigenvectors, satisfying the eigenvalue equations $\langle\psi_n^L|H{=}E_n\langle\psi_n^L|$, $H|\psi_n^R\rangle{=}E_n|\psi_n^R\rangle$, and $\langle\psi_n^L|\psi_m^R\rangle{=}\delta_{nm}$.

\begin{figure}[t]
\centering
\includegraphics[width=2.1in]{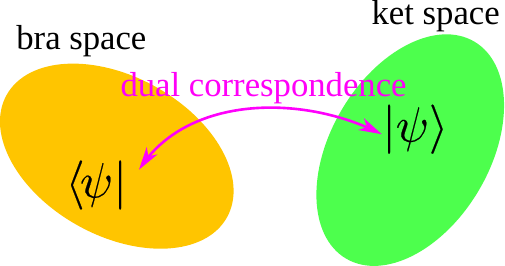} \hspace{1.7cm}
\includegraphics[width=2.1in]{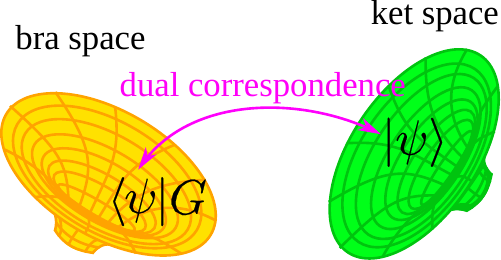}
\caption{The schematic pictures of the geometry of Hilbert space for (Left) Hermitian case and (Right) non-Hermitian case.}\label{fig:metric}
\end{figure}

Another perspective to the biorthogonal description of non-Hermitian quantum mechanics~\cite{Brody_2013} is to consider the Hilbert space geometry with a metric operator $G$~\cite{thermodynamics,DaJian2019b,Ju2019}. As the schematic pictures shown in Fig.~\ref{fig:metric}, 
the corresponding dual vector of $|\varphi\rangle$ becomes $\langle\varphi|G$, and the inner-product of two ket vectors becomes $\langle\varphi|G|\phi\rangle$. 
The connection-compatible metric operator $G$ is Hermitian, positive-definite, and satisfies the metric equation of motion, $\frac{\partial}{\partial t}G=i(GH-H^\dagger G)$. When the Hamiltonian is diagonalizable, i.e. the eigenvectors are complete, the solution of the metric equation of motion can be chosen as $G=\sum\limits_n|\psi_n^L\rangle\langle\psi_n^L|$. The completeness relation is 
\begin{equation}\label{eq:complet}
\openone=\sum_n|\psi_n^R\rangle\langle\psi_n^L|=\sum_n|\psi_n^R\rangle\langle\psi_n^R|G.
\end{equation}
In this perspective, when two or more eigenvectors coalesce at the exceptional point, i.e. the eigenvectors do not form a complete set, the proper metric should be obtained by directly solving the metric equation of motion.

Besides several topological features that share similar properties as Hermitian counterparts, 
there are unique features such as skin effects~\cite{ZWang-PRL2018}
and the exception points which do not exist in the Hermitian systems.
The former property shows a colossal boundary states and the sensitivity of the energy spectrum with respect to the boundary conditions. The latter property indicates the defectiveness of the non-Hermitian systems where the Hamiltonian is not diagonalizable at the particular parameter $\lambda_\mathrm{EP}$.
The exceptional points (EPs) of the non-Hermitian Hamiltonian are the special parameter points where both the eigenvalues and eigenvectors ``degenerate''\footnote{The more appropriate word for this kind of ``degeneracy'' is ``coalescence''.} into only one value and vector~\cite{Heiss_2012,review2019EP}. 
Suppose that $\lambda$ is the tuning parameter of the non-Hermitian Hamiltonian $H(\lambda)$. Since in general the different eigenvectors are not orthogonal with each other using the conventional inner-product, $\langle\psi_n^R(\lambda)|\psi_m^R(\lambda)\rangle\neq0$, two or more eigenvectors close to each other as $\lambda$ varies. Eventually these eigenvectors coalesce into one at the particular parameter point $\lambda_\mathrm{EP}$ called the EP. Therefore, the Hamiltonian at the EP has fewer eigenvectors, and these eigenvectors cannot span the entire Hilbert space. In contrast to the quantum critical point, the EP does not require a system that is in the thermodynamic limit.

EPs exhibit novel properties in both classical and quantum non-Hermitian systems~\cite{doppler2016dynamically,Xu2016, nature2017EP,Chen2017, Yoshida, 
Kawabata2019L,EP:coldatom, Peng2016,Zhou2018, PRXQuantum_2021,magnondevice_Sci2019}. 
Among all non-Hermitian systems, parity-time (PT)-symmetric non-Hermitian systems have real energy spectra~\cite{Bender1998, Bender2002, Bender2004, Bender2007} and have been experimentally observed in various physical systems, such as optical waveguides and microwave resonators designed with photon gain and loss at the edges, where significant sensitivity enhancement at the EPs has been observed~\cite{nature2017EP,Chen2017}.
The EPs in PT-symmetric systems correspond to the transition from PT-unbroken to PT-broken states, resulting in a conversion from real to complex energy spectra. The study of phases and phase transitions in PT-symmetric non-Hermitian Hamiltonians is of interest, and only a few references are listed here.~\cite{Ashida2017,ozturk2021observation, Gong2018,Kawabata2019X, QPT-XY,Longhi_2019F,Sun2022EPL,Banerjee_2021, PYChang2020,Pires-PRB2021, tu2021renyi,scar, wang2022non}.

Although the extensions in non-Hermitian systems have been studied by the use of fidelity, there are several different definitions of fidelity [see Table~\ref{tab:table1}] with different properties. These distinct properties make the diagnosis of non-Hermitian quantum systems obscure.
To reconcile this issue, we focus on the metricized fidelity and fidelity susceptibility defined in Ref.~\cite{Tzeng2021} 
which give several universal properties in PT symmetric non-Hermitian quantum systems.
These universal properties allow us to identify EPs and the quantum critical points in non-Hermitian quantum systems with the PT symmetry.
First, we prove the real part of the fidelity susceptibility is negative infinity at the exceptional points (EPs) for a certain condition. 
Second, we demonstrate the real part of the fidelity is always one-half if the EP is second order, where only two eigenstates and eigenenergies participate in the coalescence.
These properties allow us to identify EPs in the non-Hermitian quantum systems.
In addition, we find that for a Hermitian system near the quantum phase transition point, if we add a non-Hermitian perturbation in the Hamiltonian,
the fidelity susceptibility gets a strong enhancement from the non-Hermitian perturbation. This property allow a huge application of detecting weak quantum phase transition of the Hermitian system
by adding a non-Hermitian perturbation.

We demonstrate these general properties in two examples: the non-Hermitian two-legged Su-Schrieffer-Heeger (SSH) model and
the non-Hermitian spin-$1/2$ XXZ chain. Due to the non-interacting property for the former case, analytics expressions of the fidelity and the fidelity susceptibility can be obtained 
and we can have a direct comparison with the known phase diagram from the exact diagonalization of the Hamiltonian.
The latter case is an interacting quantum system and we demonstrate that all the general properties are hold.

We briefly summarize our new results and give the outline of the paper.
\begin{itemize}
  \item The fidelity is always real for the PT-unbroken state.
  \item The real part of the fidelity susceptibility is negative infinity for a certain condition near the EPs.
  \item The real part of the fidelity across the phase boundary of PT-unbroken and PT-broken states is one-half when the phase boundary is a second order EP.
\end{itemize}
These general properties of fidelity allow numerical computations to extract useful information about EPs by obtaining only ground-state eigenvectors rather than many excited states. Since many numerical methods are easier to compute only one eigenvector than many excited states, the fidelity with these general properties becomes a good tool for studying non-Hermitian quantum systems.

\begin{table*}
  \caption{List of different definitions of the fidelity in non-Hermitian quantum systems.
$|\psi_0^L(\lambda)\rangle$ and $|\psi_0^R(\lambda)\rangle$ are the left and right eigenvectors with the eigenvalue $E_0(\lambda)$ of the non-Hermitian Hamiltonian, $H^\dagger(\lambda)\neq H(\lambda)$, respectively. The checkmark labels the fidelity with general properties discussed in this paper. The range and the time-dependency are also listed for comparison.}\label{tab:table1}
  \begin{tabularx}{\linewidth}{clccc}
    \hline\hline
    & Formula & Range & Time-Dependency & Reference\\
    \hline
       $\checkmark$ & $\mathcal{F}=\langle\psi_0^L(\lambda)|\psi_0^R(\lambda{+}\epsilon)\rangle\langle\psi_0^L(\lambda{+}\epsilon)|\psi_0^R(\lambda)\rangle$
& $\mathbb{C}$ & independent & \cite{Tzeng2021}\\
   & $\mathcal{F}^{RR}=\langle\psi_0^R(\lambda)|\psi_0^R(\lambda{+}\epsilon)\rangle\langle\psi_0^R(\lambda{+}\epsilon)|\psi_0^R(\lambda)\rangle$
& $[0,1]$  & dependent & \cite{fsus-RR_PRL_2020}\\
   & $\mathcal{F}^{LR}_\text{half-sum}=\frac{1}{2}|\langle\psi_0^L(\lambda)|\psi_0^R(\lambda{+}\epsilon)\rangle+\langle\psi_0^R(\lambda)|\psi_0^L(\lambda{+}\epsilon)\rangle|$
& $\mathbb{R}^+$ & dependent & \cite{half} \\
   & $\mathcal{F}^{LR}_\text{sqrt-abs}=\sqrt{|\langle\psi_0^L(\lambda)|\psi_0^R(\lambda{+}\epsilon)\rangle\langle\psi_0^L(\lambda{+}\epsilon)|\psi_0^R(\lambda)\rangle|}$
& $\mathbb{R}^+$ & independent & \cite{DaJian2019a} \\
   & $\mathcal{F}^{LR}_\text{sqrt}=\sqrt{\langle\psi_0^L(\lambda)|\psi_0^R(\lambda{+}\epsilon)\rangle\langle\psi_0^L(\lambda{+}\epsilon)|\psi_0^R(\lambda)\rangle}$
& $\mathbb{C}$  & independent & \cite{fsus:perturbation}\\
    \hline\hline
  \end{tabularx}
\end{table*}

Due to the eigenvectors being labelled with left and right, many different definitions of fidelity were proposed~\cite{Tzeng2021,fsus-RR_PRL_2020,half,DaJian2019a,fsus:perturbation,Chen_2019,nishiyama2020imaginary,nishiyama2020fidelity,fsus-RR_EPL_2020,Macri:2021}. 
We list the different definitions of the fidelity in the existing literatures in Tab.~\ref{tab:table1}. 
In this paper, we focus on the metricized fidelity~\cite{Tzeng2021} which has universal properties near the quantum critical point and EPs in contrast to other proposals.
The metricized fidelity is
\begin{align}
\mathcal{F}(\lambda)&=\langle\psi_0^L(\lambda)|\psi_0^R(\lambda+\epsilon)\rangle\langle\psi_0^L(\lambda+\epsilon)|\psi_0^R(\lambda)\rangle \label{eq:fLR}\\
&=1-\mathcal{X}_F\epsilon^2+O(\epsilon^3),\label{eq:fsus}
\end{align}
where $\mathcal{X}_F$ in Eq.~\eqref{eq:fsus} is the fidelity susceptibility.
The normalization conditions are chosen to be $\langle\psi_n^L|\psi_m^R\rangle=\delta_{nm}$ and $\langle\psi_n^R|\psi_n^R\rangle=1$.

Note that the fidelity Eq.~(\ref{eq:fLR}) between two eigenstates is time-independent, and a conjecture has been made that the fidelity susceptibility Eq.~\eqref{eq:fsus} diverges to the negative infinity when the parameter $\lambda$ approaches $\lambda_{\rm EP}$~\cite{Tzeng2021}. 
For the other definitions of fidelity, e.g., the overlap between right eigenvectors only~\cite{fsus-RR_PRL_2020,fsus-RR_EPL_2020}, the fidelity sits in the region $\mathcal{F}^{RR}\in[0,1]$. However, the fidelity defined in Eq.~\eqref{eq:fLR} and the corresponding susceptibility Eq.~\eqref{eq:fsus} can be complex in general.
From the geometric perspective, the metricized fidelity compares both the wavefunctions and the metric operators $G$ for the nearby parameters. Therefore, the artificial restrictions by taking the absolute value~\cite{DaJian2019a} or the square root~\cite{fsus:perturbation} are unnecessary.

We organize the paper as follows: 
In Sec.~\ref{sec:fsus}, we prove the general properties of the fidelity and the fidelity susceptibility from the perturbation theory with a PT constrain. We demonstrate these general properties for the non-Hermitian two-legged SSH model in Sec.~\ref{sec:SSH}
and spin-$1/2$ XXZ chain with the staggered imaginary magnetic field in $z$-direction in Sec.~\ref{sec:XXZ}.
We conclude our results and discuss some future directions in Sec.~\ref{sec:conclusion}.

\section{Perturbation theory and Parity-Time constraint}\label{sec:fsus}
The derivation of fidelity and its susceptibility by the perturbation theory~\cite{You2007,fsus-RR_PRL_2020,fsus:perturbation,perturbation1972,intrinsic,perturbation_JPCM_2021} is briefly reviewed for further discussions. 
Consider a general Hamiltonian $H(\lambda)=H_0+\lambda V$, where $\lambda\in\mathbb{R}$ is a tuning parameter. 
With a small change of the parameter $\lambda$; $\lambda\to\lambda'{=}\lambda{+}\epsilon$, where $\epsilon\ll1$, then $H(\lambda{+}\epsilon)=H_0{+}(\lambda{+}\epsilon)V=H(\lambda){+}\epsilon V$. Since $\epsilon$ is a small quantity, $\epsilon V$ can be considered as a perturbation to $H(\lambda)$. Using this fact, eigenstates of $H(\lambda{+}\epsilon)$ can be expressed in terms of the eigenstates of $H(\lambda)$. Before applying the perturbation theory, let us find an expression of the fidelity susceptibility in terms of the wave functions first. The right and left eigenstates of the $H(\lambda+\epsilon)$ are expressed in the Taylor series expansion,
$|\psi_0^R(\lambda{+}\epsilon)\rangle = |\psi_0^R(\lambda)\rangle + \epsilon |\partial_\lambda\psi_0^R(\lambda)\rangle
+ \frac{\epsilon^2}{2}|\partial^2_\lambda\psi_0^R(\lambda)\rangle + O(\epsilon^3)$ and 
$\langle\psi_0^L(\lambda{+}\epsilon)| = \langle\psi_0^L(\lambda)| + \epsilon\langle\partial_\lambda\psi_0^L(\lambda)|
+ \frac{\epsilon^2}{2}\langle\partial^2_\lambda\psi_0^L(\lambda)| + O(\epsilon^3)$, respectively. Substituting this expression into the definition of the fidelity Eq.~\eqref{eq:fLR}, the following expression of the $\mathcal{X}_F(\lambda)$ is obtained, 
\begin{align}\label{eq:FS}
\mathcal{X}_F(\lambda) &= \langle\partial_\lambda\psi_0^L(\lambda)|\partial_\lambda\psi_0^R(\lambda)\rangle -\langle\psi_0^L(\lambda)|\partial_\lambda\psi_0^R(\lambda)\rangle \langle\partial_\lambda\psi_0^L(\lambda)|\psi_0^R(\lambda)\rangle,
\end{align}
where $\partial_\lambda\langle\psi_0^L(\lambda)|\psi_0^R(\lambda)\rangle=0$ 
and $\partial^2_\lambda\langle\psi_0^L(\lambda)|\psi_0^R(\lambda)\rangle{=}0$ have been used in the above derivation.

Now, let us use the perturbation theory for the expression of Eq.~\eqref{eq:FS} which does not involve differentiation. Following the conventional perturbation approach, $|\psi_0^R(\lambda{+}\epsilon)\rangle$, $\langle\psi_0^L(\lambda{+}\epsilon)|$, and $E_0(\lambda{+}\epsilon)$ which are right, left eigen-states and eigen-energy of the $H(\lambda+\epsilon)$ respectively can be written as follows:
\begin{align}
&|\psi_0^R(\lambda{+}\epsilon)\rangle=|\psi_0^R(\lambda)\rangle+\epsilon|\psi_0^{R,(1)}(\lambda)\rangle +O(\epsilon^2),\label{eq:pertur_R}\\
&\langle \psi_0^L(\lambda{+}\epsilon)|=\langle \psi_0^L(\lambda)|+\epsilon\langle \psi_0^{L,(1)}(\lambda)|+O(\epsilon^2),\label{eq:pertur_L}\\
&E_0(\lambda{+}\epsilon)=E_0(\lambda)+\epsilon E_0^{(1)}(\lambda) +\epsilon^2E_0^{(2)}(\lambda) + O(\epsilon^3),
\end{align}
where the superscript number $(i)$ denotes the order of $\epsilon$. Substituting the above equations to the eigenvalue equations $H(\lambda{+}\epsilon)|\psi_0^R(\lambda{+}\epsilon)\rangle=E_0(\lambda{+}\epsilon)|\psi_0^R(\lambda{+}\epsilon)\rangle$ and $\langle\psi_0^L(\lambda{+}\epsilon)|H(\lambda{+}\epsilon)=\langle\psi_0^L(\lambda{+}\epsilon)|E_0(\lambda{+}\epsilon)$ gives the following identities in the first order of $\epsilon$,
\begin{align}
	&H(\lambda)|\psi_0^{R,(1)}(\lambda)\rangle + V|\psi_0^R(\lambda)\rangle =E_0(\lambda)|\psi_0^{R,(1)}(\lambda)\rangle+E_0^{(1)}(\lambda)|\psi_0^R(\lambda)\rangle,\label{eq:firstOrderEq1}\\
	&\langle \psi_0^{L,(1)}(\lambda)|H(\lambda)+\langle \psi_0^L(\lambda)|V =\langle \psi_0^{L,(1)}(\lambda)|E_0(\lambda)+\langle \psi_0^L(\lambda)|E_0^{(1)}(\lambda).\label{eq:firstOrderEq2}
\end{align}
By applying $\langle \psi_0^L(\lambda)|$ and $|\psi_0^R(\lambda)\rangle$ to the Eqs.\eqref{eq:firstOrderEq1} and \eqref{eq:firstOrderEq2} respectively, we get the first-order energy correction $E_0^{(1)}(\lambda)=\langle \psi_0^L(\lambda) |V|\psi_0^R(\lambda)\rangle $. Applying the $\langle \psi_n^L(\lambda)|$ and $|\psi_n^R(\lambda)\rangle$ with non-zero $n\neq0$ to each equation gives:
\begin{align}
	&\langle \psi_n^L(\lambda)|\psi_0^{R,(1)}(\lambda)\rangle=\frac{\langle \psi_n^L(\lambda)|V|\psi_0^R(\lambda)\rangle }{E_0(\lambda)-E_n(\lambda)},\\	
	&\langle \psi_0^{L,(1)}(\lambda)|\psi_0^{R}(\lambda)\rangle=\frac{\langle \psi_0^L(\lambda)|V|\psi_n^R(\lambda)\rangle }{E_0(\lambda)-E_n(\lambda)}.
\end{align}
Additionally, from Eqs.~\eqref{eq:pertur_R} and \eqref{eq:pertur_L} with the normalization condition, $\langle\psi_0^L(\lambda{+}\epsilon)|\psi_0^R(\lambda{+}\epsilon)\rangle{=}1$, we have
$\langle\psi_0^L(\lambda)|\psi_0^{R,(1)}(\lambda)\rangle+\langle\psi_0^{L,(1)}(\lambda)|\psi_0^R(\lambda)\rangle=0$.
As a result, up to the first order of $\epsilon$,
\begin{align*}
	|\psi_0^R(\lambda+\epsilon)\rangle
	&=\Big(1+\epsilon\langle \psi_0^L(\lambda)|\psi_0^{R,(1)}(\lambda)\rangle \Big)|\psi_0^R(\lambda)\rangle +\epsilon\sum_{n\neq 0}\frac{\langle \psi_n^L(\lambda)|V|\psi_0^R(\lambda)\rangle }{E_0(\lambda)-E_n(\lambda)} |\psi_n^R(\lambda)\rangle +O(\epsilon^2),\\
	\langle \psi_0^L(\lambda+\epsilon)|
	&=\Big(1+\epsilon\langle \psi_0^{L,(1)}(\lambda)|\psi_0^{R}(\lambda)\rangle \Big)\langle \psi_0^L(\lambda)|+\epsilon\sum_{n\neq 0}\frac{\langle \psi_0^L(\lambda)|V|\psi_n^R(\lambda)\rangle }{E_0(\lambda)-E_n(\lambda)} \langle \psi_n^L(\lambda)| +O(\epsilon^2),
\end{align*}
where the completeness Eq.~\eqref{eq:complet} 
has been used. 
Here note that terms proportional to $\epsilon\langle \psi_0^L(\lambda)|\psi_0^{R,(1)}(\lambda)\rangle$ and $\epsilon\langle \psi_0^{L,(1)}(\lambda)|\psi_0^{R}(\lambda)\rangle$ are related to the change of the $U(1)$ phase factors of the states or Berry phases. These phase factors turn out to disappear in the final expression of the $\mathcal{X}_F(\lambda)$ as we will see soon. From the above results, we can get alternative expressions of $|\partial_\lambda\psi_0^{R}(\lambda)\rangle$ and $\langle \partial_\lambda \psi^{L}_0(\lambda)|$ as follows,
\begin{align*}
	|\partial_\lambda \psi_0^R(\lambda)\rangle&=\lim_{\epsilon\rightarrow 0}\frac{|\psi_0^R(\lambda+\epsilon)\rangle-|\psi_0^R(\lambda)\rangle}{\epsilon} \\
	&=\langle \psi_0^L(\lambda)|\psi_0^{R,(1)}(\lambda)\rangle |\psi_0^R(\lambda)\rangle +\sum_{n\neq 0}\frac{\langle \psi_n^L(\lambda)|V|\psi_0^R(\lambda)\rangle }{E_0(\lambda)-E_n(\lambda)} |\psi_n^R(\lambda)\rangle,\\
	\langle \partial_\lambda \psi_0^L(\lambda)|&=\lim_{\epsilon\rightarrow 0}\frac{\langle \psi_0^L(\lambda+\epsilon)|-\langle \psi_0^L(\lambda)|}{\epsilon} \\
	&=\langle \psi_0^{L,(1)}(\lambda)|\psi_0^{R}(\lambda)\rangle \langle \psi_0^L(\lambda)| +\sum_{n\neq 0}\frac{\langle \psi_0^L(\lambda)|V|\psi_n^R(\lambda)\rangle }{E_0(\lambda)-E_n(\lambda)} \langle \psi_n^L(\lambda)|.
\end{align*}
Substituting these results to Eq.\eqref{eq:FS}, the fidelity susceptibility is
\begin{align}\label{eq:chiF}
	\mathcal{X}_F(\lambda)=\sum_{n\neq0}\frac{\langle\psi_0^L(\lambda)|V|\psi_n^R(\lambda)\rangle \langle\psi_n^L(\lambda)|V|\psi_0^R(\lambda)\rangle}{[E_0(\lambda)-E_n(\lambda)]^2}.
\end{align}
Here as it is mentioned before, the terms related to the U(1) phase factors disappear in the final expressions.

Moreover, applying the perturbation theory up to second order of $\epsilon$ gives the following expression of the second order energy correction $E_0^{(2)}$, which is proportional to the second derivatives of the ground state energy $\partial_\lambda^2E_0(\lambda)$~\cite{perturbation1972},
\begin{align}
 E_0^{(2)}(\lambda)=\sum_{n\neq0}\frac{\langle\psi_0^L(\lambda)|V|\psi_n^R(\lambda)\rangle \langle\psi_n^L(\lambda)|V|\psi_0^R(\lambda)\rangle}{E_0(\lambda)-E_n(\lambda)}.
\end{align}
Compare with the right-right fidelity susceptibility~\cite{fsus-RR_PRL_2020},
\begin{align}
\mathcal{X}_F^{RR}(\lambda)=&\sum_{m,n\neq0}
\frac{\langle\psi_0^R(\lambda)|V^\dagger |\psi_m^L(\lambda)\rangle}{[E_0(\lambda)-E_m(\lambda)]^*}
\frac{\langle\psi_n^L(\lambda)|V|\psi_0^R(\lambda)\rangle}{E_0(\lambda)-E_n(\lambda)}  
\Big(\langle\psi_m^R(\lambda)|\psi_n^R(\lambda)\rangle-\langle\psi_m^R(\lambda)|\psi_0^R(\lambda)\rangle \langle\psi_0^R(\lambda)|\psi_n^R(\lambda)\rangle\Big),
\end{align}
the metricized fidelity susceptibility Eq.~\eqref{eq:chiF} is more similar to the second derivative of the ground state energy.
However, unlike the right-right fidelity $\mathcal{F}^{RR}(\lambda)=\abs{\langle\psi_0^R(\lambda)|\psi_0^R(\lambda+\epsilon)\rangle}^2$ which is guaranteed to be a real value between 0 and 1 by the self-normalization, the metricized fidelity and its susceptibility are complex in general. It is interesting to see the consequence if the condition of PT-symmetry is imposed, as discussed in the following subsection.

\subsection{Constraints by the PT-symmetry}\label{sec:constraint}
Based on the results of the perturbation theory in the previous subsection,
we derive the constraints on the fidelity susceptibility by the PT-symmetry. Suppose a general non-Hermitian Hamiltonian is given by $H(\lambda)=H_0+\lambda V$ where $\lambda$ is a real parameter. Since $H(\lambda)$ preserves the PT-symmetry for any $\lambda\in\mathbb{R}$, $H_0$ and $V$ preserves the PT-symmetry independently; $\mathcal{PT}H_0(\mathcal{PT})^{-1}=H_0$ and $\mathcal{PT}V(\mathcal{PT})^{-1}=V$. Here we consider the time-reversal operation is just given by a complex conjugate $\mathcal{K}$ and satisfying $\mathcal{T}^2=1$.\footnote{For the anti-PT symmetric case, i.e. $\mathcal{T}^2=-1$ or $(\mathcal{PT})^2=-1$, the usual perturbation theory is not applicable due to the Kramer's theorem. In this paper we only focus on the PT symmetric case.} As a result, the operator $\mathcal{PT}$ is an anti-unitary operator satisfying following properties,
\begin{subequations}\label{eq:propertiesOfPToperator}
\begin{align}
	\mathcal{PT}c|\psi\rangle= c^*\mathcal{PT}|\psi\rangle,\\
	\langle \psi|\phi\rangle=\langle \mathcal{PT}\phi|\mathcal{PT}\psi\rangle,\\
	(\mathcal{PT})^2=1.
\end{align}
\end{subequations}
These properties of the PT operator give constraints on the structures of the eigenstates and eigenvalues of the Hamiltonian $H(\lambda)$.

The eigenvalue equations are
$\langle \psi_n^L|H(\lambda)=\langle \psi_n^L|E_n$, and $H(\lambda)|\psi_n^R\rangle=E_n|\psi_n^R\rangle$.
Due to non-Hermiticity of the $H(\lambda)$, the eigenvalue $E_n$ is generally a complex number. 
Depending on whether the eigenvalue $E_n$ has a nonzero imaginary part, 
the constraints by the $\mathcal{PT}$ symmetry turn out to be different. 

First, let us consider the case that the eigenvalue $E_n$ is given by a real number. From the facts that $E_n=E_n^*$, $[H(\lambda),\mathcal{PT}]=0$ and properties of the PT operator, Eqs.~\eqref{eq:propertiesOfPToperator}, one can easily prove that $\mathcal{PT}|\psi_n^R\rangle$ and $(\mathcal{PT}|\psi_n^L\rangle)^\dagger$ are also eigenstates of the $H(\lambda)$ with the same eigenvalue $E_n$.
Since $(\mathcal{PT})^2=1$, there is no Kramer degeneracy.
Therefore it is assumed that there is no protected degeneracy generally. Then, $\mathcal{PT}|\psi_n^R\rangle$ and $(\mathcal{PT}|\psi_n^L\rangle)^\dagger$ should be identical to the $|\psi_n^R\rangle$ and $\langle \psi_n^L|$ up to phase factors, respectively. It means that both $\langle \psi_n^L|$ and $|\psi_n^R\rangle$ are the eigenstates of the $\mathcal{PT}$ operator; $\mathcal{PT}|\psi_n^R\rangle=e^{i\theta_{n}^R}|\psi_n^R\rangle$, $\mathcal{PT}|\psi_n^L\rangle=e^{i\theta_n^L}|\psi_n^L\rangle$. Additionally, from $\langle \psi_n^L|\psi_n^R\rangle=\langle \mathcal{PT}\psi_n^R|\mathcal{PT}\psi_n^L\rangle=1$, it shows that $\theta_n^L=\theta_n^R\equiv \theta_n$, which means that $|\psi_n^R\rangle$ and $|\psi_n^L\rangle$ are degenerated eigenstates of the $\mathcal{PT}$ operator. 

Now let us consider the case when $E_n$ is complex, i.e. $E_n^*\neq E_n$. The energy eigenstates $\langle \psi_n^L|$ and $|\psi_n^R\rangle$ are not the eigenstates of the $\mathcal{PT}$ operator anymore. Instead, $(\mathcal{PT}|\psi_n^L\rangle)^\dagger$ and $\mathcal{PT}|\psi_n^R\rangle$ are the other eigenstates of $H(\lambda)$ with the energy eigenvalue $E_n^*$. Therefore there are always pairs of left and right eigenstates with their PT-symmetry partners.

Finally, based on the above facts, a general structure of the eigenstates and eigenvalues of the Hamiltonian $H(\lambda)$ which has $\mathcal{PT}$ symmetry can be obtained. Suppose the $H(\lambda)$ has total $N$ eigenvalues without degeneracy. Among them, $N_r$ energy eigenvalues are real while $N-N_r\equiv N_c$ eigenvalues are complex energy eigenvalues with nonzero imaginary parts, where $N_c$ is even. Then the structures of the eigenstates and eigenvalues of $H(\lambda)$ are given as follows:

\begin{subequations}\label{eq:structureHamiltonianPTsymmetry}
For the real energy eigenvalues, $E_{r,n}^*=E_{r,n}$,  $n=1,\cdots,N_r$,
\begin{align}
	&H(\lambda)|\psi_{r,n}^R\rangle=E_{r,n}|\psi_{r,n}^R\rangle,\\
	&\langle \psi_{r,n}^L|H(\lambda)=\langle \psi_{r,n}^L|E_{r,n},\nonumber\\
	&\mathcal{PT}|\psi_{r,n}^{R/L}\rangle=e^{i\theta_n}|\psi_{r,n}^{R/L}\rangle.\;\nonumber
\end{align}

For the complex energy eigenvalues, $E_{c,n}^* \neq E_{c,n}$,  $n=1,\cdots,N_c /2$,
\begin{align}
	&\begin{cases} 
		H(\lambda) |\psi_{c,n}^R\rangle=E_{c,n}|\psi^R_{c,n}\rangle \\ H(\lambda)|\bar{\psi}^R_{c,n}\rangle=E_{c,n}^*|\bar{\psi}^R_{c,n}\rangle 
	\end{cases},
	\begin{cases}
		\langle \psi_n^L|H(\lambda)=\langle \psi_n^L(\lambda)|E_{c,n}\\
		\langle \bar{\psi}_n^L|H(\lambda)=\langle \bar{\psi}_n^L(\lambda)|E^*_{c,n}
	\end{cases}\\
	&|\bar{\psi}_{c,n}^{R/L}\rangle=\mathcal{PT}|\psi_{c,n}^{R/L}\rangle,\; \psi_{c,n}^{R/L}=\mathcal{PT}|\bar{\psi}_{c,n}^{R/L}\rangle.\nonumber
\end{align}

The orthonormality and completeness,
\begin{align}
	&\begin{cases}
	    \langle\psi_{r,n}^L|\psi_{r,m}^R\rangle=\langle \psi_{c,n}^L|\psi_{c,m}^R\rangle=\langle \bar{\psi}_{c,n}^L|\bar{\psi}_{c,m}^R\rangle=\delta_{nm}\\
		\langle\psi_{r,n}^L|\psi_{c,m}^R\rangle=\langle\psi_{r,n}^L|\bar{\psi}_{c,m}^R\rangle=\langle \psi_{c,n}^L|\psi_{r,m}^R\rangle=0\\
		\langle \psi_{c,n}^L|\bar{\psi}_{c,m}^R\rangle=\langle \bar{\psi}_{c,n}^L|\psi_{r,m}^R\rangle =\langle \bar{\psi}_{c,n}^L|\psi_{c,m}^R\rangle=0
	\end{cases}\\
	&\sum_{n=1}^{N_r}|\psi_{r,n}^R\rangle\langle \psi_{r,n}^L|+\sum_{n=1}^{N_c/2}\Big(|\psi_{c,n}^R\rangle\langle\psi_{c,n}^L|+|\bar{\psi}_{c,n}^R\rangle\langle\bar{\psi}_{c,n}^L|\Big)=\openone,\nonumber
\end{align}
\end{subequations}
where the subscript $r$ and $c$ denote `real' and `complex' respectively. Note that number of complex eigenvalues $N_c$ is always even due to the PT symmetry. 

Now using the Eqs.~\eqref{eq:structureHamiltonianPTsymmetry} and Eq.~\eqref{eq:chiF}, let us derive constraints on the fidelity susceptibility by the PT-symmetry. There are two cases:
\begin{enumerate}
\item[(i)] The left and right ground states are the PT-unbroken states ($E_0^*{=}E_0$), and 
\item[(ii)] The left and right ground states are the PT-broken states ($E_0^*{\neq}E_0$).
\end{enumerate}

In the case (i), it is easy to prove that the fidelity is real.
\begin{proof}
$\mathcal{F}^*=\langle\psi_0^R(\lambda+\epsilon)|\psi_0^L(\lambda)\rangle\langle\psi_0^R(\lambda)|\psi_0^L(\lambda+\epsilon)\rangle
= \langle\mathcal{PT}\psi_0^L(\lambda)|\mathcal{PT}\psi_0^R(\lambda+\epsilon)\rangle\langle\mathcal{PT}\psi_0^L(\lambda+\epsilon)|\mathcal{PT}\psi_0^R(\lambda)\rangle
= \langle\psi_0^L(\lambda)|\psi_0^R(\lambda+\epsilon)\rangle\langle\psi_0^L(\lambda+\epsilon)|\psi_0^R(\lambda)\rangle
=\mathcal{F}$. Since the fidelity is real, the fidelity susceptibility $\mathcal{X}_F$ is also real.
\end{proof}

In the case (ii) PT-broken states, the fidelity susceptibility is given as follows:
\begin{align}
\mathcal{X}_F=&\frac{\langle \psi_0^L|V|\bar{\psi}_0^R\rangle\langle \bar{\psi}_{0}^L|V|\psi_0^R\rangle}{[E_0-E_0^*]^2}
+\sum_{n=1}^{N_r}\frac{\langle \psi_0^L|V|\psi_{r,n}^R\rangle\langle \psi_{r,n}^L|V|\psi_0^R\rangle }{[E_0-E_{r,n}]^2} \nonumber\\
&+\sum_{n=2}^{N_c/2}\Bigg[\frac{\langle \psi_0^L|V|\psi_{c,n}^R\rangle\langle\psi_{c,n}^L|V|\psi_0^R\rangle}{[E_0-E_{c,n}]^2}
+\frac{\langle \psi_0^L|V|\bar{\psi}_{c,n}^R\rangle\langle\bar{\psi}_{c,n}^L|V|\psi_0^R\rangle}{[E_0-E^*_{c,n}]^2}\Bigg],
\end{align}
where $|\psi_{0}^{R/L}\rangle=|\psi_{c,n=1}^{R/L}\rangle$, $|\bar{\psi}_0^{R/L}\rangle=\mathcal{PT}|\psi_0^{R/L}\rangle$ and $E_0=E_{c,n=1}$. 
The following identity of $\mathcal{X}_F$ in the case of the PT-broken states can be obtained:
\begin{align}
\mathcal{X}_F=&\frac{|\langle \psi_0^L|V|\bar{\psi}_0^R\rangle|^2}{[E_0-E_0^*]^2}+\sum_{n=1}^{N_r}\frac{\langle \psi_{r,n}^R|V^\dagger |\bar{\psi}_0^L\rangle\langle \bar{\psi}_0^R|V^\dagger |\psi_{r,n}^L\rangle}{[E_0-E_{r,n}]^2}
\nonumber\\
&+\sum_{n=2}^{N_c/2}\Bigg[\frac{\langle \bar{\psi}_{c,n}^R|V^\dagger |\bar{\psi}_0^L\rangle\langle \bar{\psi}_0^R|V^\dagger |\bar{\psi}_{c,n}^L\rangle}{[E_0-E_{c,n}]^2}
+\frac{\langle \psi_{c,n}^R|V^\dagger |\bar{\psi}_0^L\rangle\langle \bar{\psi}_0^R|V^\dagger |\psi_{c,n}^L\rangle}{[E_0-E_{c,n}^*]^2}\Bigg] \neq [\mathcal{X}_F]^*,
\end{align}
where the identities $\langle\bar{\psi}_0^L|V|\psi_0^R\rangle=\langle \mathcal{PT}\psi_0^R|V^\dagger|\mathcal{PT}\bar{\psi}_0^L\rangle=\langle \bar{\psi}_0^R|V^\dagger |\psi_0^L\rangle =[\langle \psi_0^L|V|\bar{\psi}_0^R\rangle]^*$ have been used for the first term. Note that the above expression is different from the $[\mathcal{X}_F]^*$. Instead, the complex conjugation of the fidelity susceptibility is
\begin{align}
	[\mathcal{X}_F]^*=&\frac{|\langle \psi_0^L|V|\bar{\psi}_0^R\rangle|^2}{[E^*_0-E_0]^2}+\sum_{n=1}^{N_r}\frac{\langle \bar{\psi}_0^L|V|\psi_{r,n}^R\rangle\langle \psi_{r,n}^L|V|\bar{\psi}_0^R\rangle }{[E_0^*-E_{r,n}]^2}\nonumber\\
	&+\sum_{n=2}^{N_c/2}\Bigg[\frac{\langle \bar{\psi}_0^L|V|\psi_{c,n}^R\rangle\langle\psi_{c,n}^L|V|\bar{\psi}_0^R\rangle}{[E_0^*-E_{c,n}]^2}
+\frac{\langle\bar{\psi}_0^L|V|\bar{\psi}_{c,n}^R\rangle\langle\bar{\psi}_{c,n}^L|V|\bar{\psi}_0^R\rangle}{[E_0^*-E^*_{c,n}]^2}\Bigg] =\bar{\mathcal{X}}_F
\end{align} 
This is nothing but the fidelity susceptibility from the PT-partner states,
\\
$\bar{\mathcal{F}}(\lambda)\equiv \langle \bar{\psi}_0^L(\lambda)|\bar{\psi}_0^R(\lambda{+}\epsilon)\rangle\langle \bar{\psi}_0^L(\lambda{+}\epsilon)|\bar{\psi}_0^R(\lambda)\rangle
\approx 1-\bar{\mathcal{X}}_F\epsilon^2$. As a result, one can define the following new fidelity susceptibility $\mathcal{X}_F^\mathrm{real}\equiv (\mathcal{X}_F+\bar{\mathcal{X}}_F)/2$ which has always real value in the PT-broken phase as follows:
\begin{align}\label{eq:realFidelitySusceptibility}
	\mathcal{X}_F^\mathrm{real} =&\frac{|\langle \psi_0^L|V|\bar{\psi}_0^R\rangle|^2}{[E_0-E_0^*]^2}
	+\frac{1}{2}\Bigg[\sum_{n=1}^{N_r}\frac{\langle \psi_{r,n}^R|V^\dagger |\bar{\psi}_0^L\rangle\langle \bar{\psi}_0^R|V^\dagger |\psi_{r,n}^L\rangle}{[E_0-E_{r,n}]^2} \nonumber\\
	&+\sum_{n=2}^{N_c/2}\Bigg(\frac{\langle \bar{\psi}_{c,n}^R|V^\dagger |\bar{\psi}_0^L\rangle\langle \bar{\psi}_0^R|V^\dagger |\bar{\psi}_{c,n}^L\rangle}{[E_0-E_{c,n}]^2}+\frac{\langle \psi_{c,n}^R|V^\dagger |\bar{\psi}_0^L\rangle\langle \bar{\psi}_0^R|V^\dagger |\psi_{c,n}^L\rangle}{[E_0-E_{c,n}^*]^2}\Bigg)+\mathrm{H.c.}\Bigg].
\end{align}
Equivalently, $\mathcal{X}_F^\mathrm{real}=\mathrm{Re}[\mathcal{X}_F]$ is the real part of the fidelity susceptibility, and it takes care of both the PT-broken states its PT-partner states.
Suppose there is an exceptional point $\lambda_\mathrm{EP}$ where two complex energies $E_0(\lambda_\mathrm{EP})$ and $E_0^*(\lambda_\mathrm{EP})$ meet.
Here the regime $\lambda>\lambda_\mathrm{EP}$ is assumed to be the PT-broken regime. As we approach to the EP from the PT-broken state ($\lambda\to\lambda_\mathrm{EP}^+$), the first term in Eq.~\eqref{eq:realFidelitySusceptibility} become dominant, and it leads to diverging negative-real valued function behavior of the susceptibility,
\begin{align}
\lim_{\lambda\to\lambda_\mathrm{EP}^+}\mathcal{X}_F^\mathrm{real}(\lambda)\approx \lim_{\lambda\to\lambda_\mathrm{EP}^+}\frac{|\langle \psi_0^L|V|\bar{\psi}_0^R\rangle|^2}{[E_0-E_0^*]^2}= -\infty,
\label{eq:xireal}
\end{align}
where the denominator $[E_0-E_0^*]^2<0$ and the numerator is not zero.\footnote{When the numerator of Eq.~(\ref{eq:xireal}) also approaches to zero, other contributions from Eq.~(\ref{eq:realFidelitySusceptibility}) need to be considered and it can lead to a positive value as shown in Fig.~\ref{fig:SSHDensityPlot}.} 
This result supports the conjecture that the real part of the fidelity susceptibility diverges to negative infinity as the parameter approaches to the EP~\cite{Tzeng2021},
\begin{equation}\label{eq:negative_infty}
\lim_{\lambda\to\lambda_\mathrm{EP}}\mathrm{Re}[\mathcal{X}_F(\lambda)]=-\infty.
\end{equation}
In this argument, we assume that the remaining terms in the real-value fidelity susceptibility Eq.~\eqref{eq:realFidelitySusceptibility} are much smaller than the first term near the EP. However if there is a critical point where the excited energies, i.e. $E_{r,n>0}$, $E_{c,n>1}$, become very close to the ground state energy $E_0$, these remaining terms can become large enough to give a significant contribution to the fidelity susceptibility, and $\mathrm{Re}[\mathcal{X}_F]$ can show different behaviors.
In Sec.\ref{sec:SSH} and~\ref{sec:XXZ}, we will see that real part of fidelity susceptibility density shows a positive peak due to the quantum criticality.

In the end of this subsection, we simply apply the above constraints to the general non-interacting two-band model.
Since the total Hamiltonian is block diagonalized into different momentum sectors and the PT-symmetry holds for each block Hamiltonian, $\mathcal{H}_k$ is a $2\times2$ matrix and $[\mathcal{H}_k(\lambda),\mathcal{PT}]=0$, the constraints by the PT-symmetry can be applied to each momentum sector separately. Each momentum sector is composed of two eigenstates of $\mathcal{H}_k(\lambda)$ and the ground state $|\psi_0^R\rangle$ is given by $|\psi_0^R(\lambda)\rangle=\prod_kd^{R,\dagger}_{-}(k)\ket{0}$, where $d_\sigma^{R,\dagger}(k)$ is the right creation operator at $k$ for the eigenstate with band index $\sigma=\pm$, and $\ket{0}$ is the vacuum state. The fidelity susceptibility of $|\psi_0^R(\lambda)\rangle$ is given by the sum of fidelity susceptibility at each momentum, $\mathcal{X}_F=\sum_{k}\chi_{k,\sigma}$, where
\begin{equation}\label{eq:bandmomentumfsus}
\chi_{k,\sigma}=\frac{\langle L_\sigma(k)|V_k|R_{-\sigma}(k)\rangle\langle L_{-\sigma}(k)|V_k|R_\sigma(k)\rangle}{[\varepsilon_\sigma(k)-\varepsilon_{-\sigma}(k)]^2},
\end{equation}
and $\varepsilon_\sigma(k)$ is the single-particle energy, $|R_{\sigma}(k)\rangle=d_\sigma^{R,\dagger}(k)\ket{0}$ is the $2\times1$ single-particle right eigenvector, and $V_k$ is a $2\times2$ matrix with PT-symmetry.
Note that $\chi_{k,\sigma}$ is a real function by the constraint from the PT-symmetry, and $\chi_{k,\sigma}=\chi_{k,-\sigma}$. Even for the PT-broken case which yields complex value generally, $\chi_{k,\sigma}$ is real and negative since there are only two states at each momentum space which are PT-partner states to each other. 

\begin{figure}[t]
\centering
\includegraphics[width=3.2in]{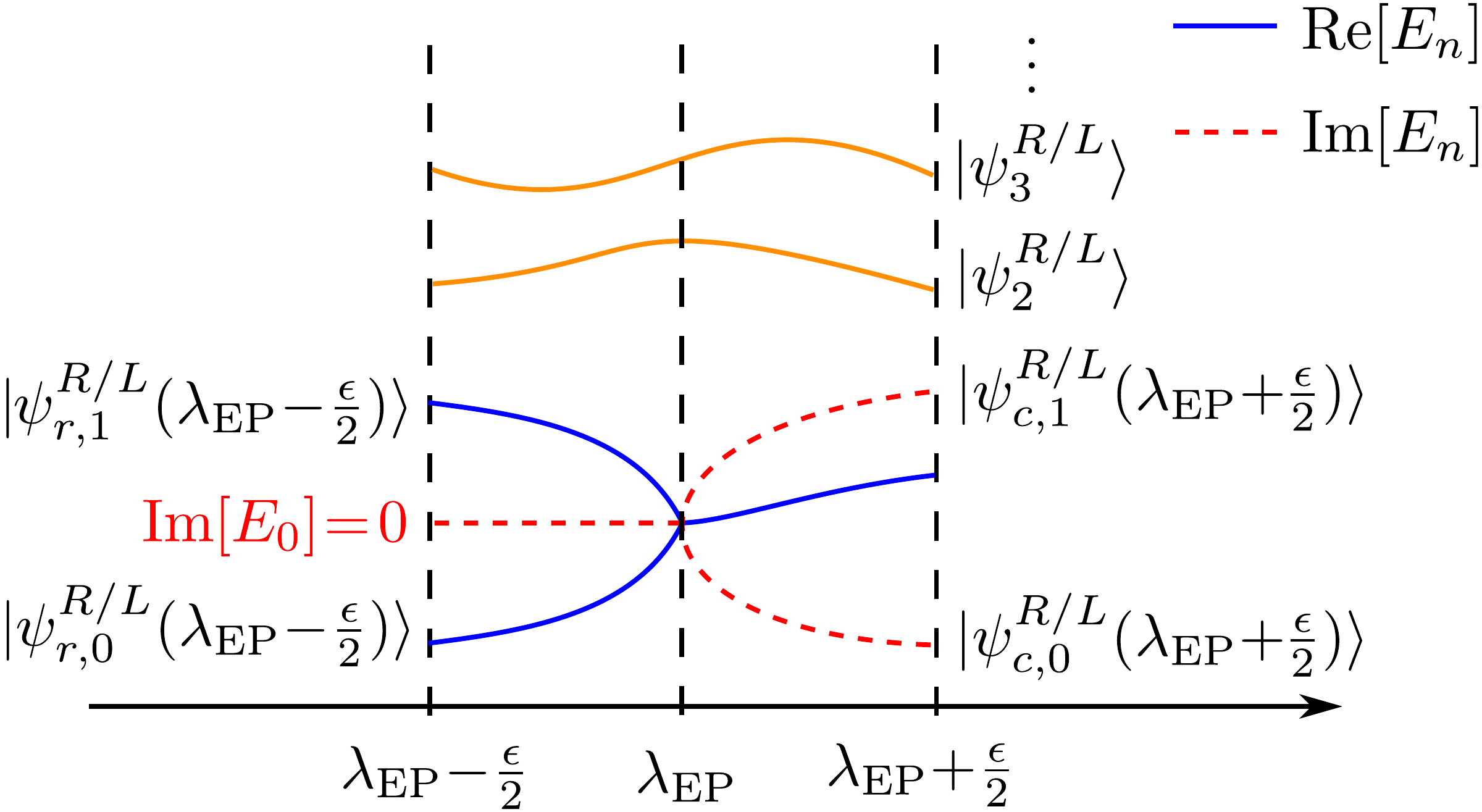}
\caption{A schematic picture for eigenstates near an exceptional point $\lambda_\mathrm{EP}$. The eigenstate $|\psi_{r,0}^{R/L}(\lambda_\mathrm{EP}{-}\frac{\epsilon}{2})\rangle$ is the PT-unbroken ground state with a real eigenvalue, and the eigenstate $|\psi_{c,0}^{R/L}(\lambda_\mathrm{EP}{+}\frac{\epsilon}{2})\rangle$ is the PT-broken ground state with a complex eigenvalue. The eigenstates $|\psi_{n}^{R/L}\rangle$ with $n>1$ are the excited states separated from the states $|\psi_{r/c,0}^{R/L}\rangle$ and $|\psi_{r/c,1}^{R/L}\rangle$ by a finite energy gap.  These excited states can be PT-broken or PT-unbroken. The PT-broken state $|\psi_{c,1}^{R/L}(\lambda_\mathrm{EP}{+}\frac{\epsilon}{2})\rangle=\mathcal{PT}|\psi_0^{R/L}(\lambda_\mathrm{EP}{+}\frac{\epsilon}{2})\rangle$ is the PT partner of the state $|\psi_{c,0}^{R/L}(\lambda_\mathrm{EP}{+}\frac{\epsilon}{2})\rangle$.}
\label{fig:OneHalfFidelity}
\end{figure}
\subsection{One-half fidelity between PT-unbroken and PT-broken states}
Before going further, here we consider an additional fact related to the fidelity itself rather than the fidelity susceptibility. Up to now, we have considered fidelity between two PT-unbroken states or between two PT-broken states. However, the fidelity can also be computed between a PT-unbroken state and a PT-broken state across an EP lying between these two states in the parameter space. 
The fidelity in this case is written as
\begin{align}
	\mathcal{F}(\lambda_{\rm EP},\epsilon)
	&\equiv \langle \psi_{r,0}^L(\lambda_{\rm EP}-\tfrac{\epsilon}{2})|\psi_{c,0}^R(\lambda_{\rm EP}+\tfrac{\epsilon}{2})\rangle\langle \psi_{c,0}^L(\lambda_{\rm EP}+\tfrac{\epsilon}{2})|\psi_{r,0}^R(\lambda_{\rm EP}-\tfrac{\epsilon}{2})\rangle,
	\label{eq:centerEP}
\end{align}
where $\epsilon$ is a positive small value and $\lambda_{\rm EP}$ is the EP where two PT-unbroken states $|\psi_{r,0}^{R/L}\rangle$ and $|\psi_{r,1}^{R/L}\rangle$ coalesce, as shown in the schematic picture Fig.~\ref{fig:OneHalfFidelity}. 
In the usual cases, the fidelity in the $\epsilon\to0$ limit is 1. However, when there is an EP, it turns out that real part of the fidelity in the $\epsilon\to0$ limit is $1/2$, and the susceptibility does not have a proper definition from Eq.~\eqref{eq:fsus}.

\begin{proof} This can be proved by using the PT-symmetry as follows:
\begin{align}
[\mathcal{F}(\lambda_{\rm EP},\epsilon)]^*&=\langle \psi_{c,0}^R(\lambda_{\rm EP}+\tfrac{\epsilon}{2})|\psi_{r,0}^L(\lambda_{\rm EP}-\tfrac{\epsilon}{2})\rangle 
\langle \psi_{r,0}^R(\lambda_{\rm EP}-\tfrac{\epsilon}{2})|\psi_{c,0}^L(\lambda_{\rm EP}+\tfrac{\epsilon}{2})\rangle\nonumber\\
	&=\langle \mathcal{PT}\psi_{r,0}^L(\lambda_{\rm EP}-\tfrac{\epsilon}{2})|\mathcal{PT}\psi_{c,0}^R(\lambda_{\rm EP}+\tfrac{\epsilon}{2})\rangle \langle \mathcal{PT}\psi_{c,0}^L(\lambda_{\rm EP}+\tfrac{\epsilon}{2})|\mathcal{PT}\psi_{r,0}^R(\lambda_{\rm EP}-\tfrac{\epsilon}{2})\rangle\nonumber\\
	&=\langle \psi_{r,0}^L(\lambda_{\rm EP}-\tfrac{\epsilon}{2})|\psi_{c,1}^R(\lambda_{\rm EP}+\tfrac{\epsilon}{2})\rangle\langle \psi_{c,1}^{L}(\lambda_{\rm EP}+\tfrac{\epsilon}{2})|\psi_{r,0}^R(\lambda_{\rm EP}-\tfrac{\epsilon}{2})\rangle.
\end{align}
Here, the fact that $\mathcal{PT}|\psi_{r,0}^{R/L}\rangle=e^{i\theta_0}|\psi_{r,0}^{R/L}\rangle$ for the PT-unbroken state is used, and the state $|\psi_{c,1}^{R/L}\rangle=\mathcal{PT}|\psi_{c,0}^{R/L}\rangle$ is the PT partner of the PT-broken state $|\psi_{c,0}^{R/L}\rangle$. 

Then the real part of the fidelity in the $\epsilon\to0$ limit is 
\begin{align}
	\lim_{\epsilon\to0} \mathrm{Re}\mathcal{F}(\lambda_{\rm EP},\epsilon)=\lim_{\epsilon\to0}\frac{\mathcal{F}(\lambda_{\rm EP},\epsilon)+[\mathcal{F}(\lambda_{\rm EP},\epsilon)]^*}{2} 
	=\frac{1}{2}\lim_{\epsilon\to0}\langle \psi_{r,0}^L(\lambda_{\rm EP}-\tfrac{\epsilon}{2})|\mathbb{P}_{\lambda_{\rm EP}+\frac{\epsilon}{2}}|\psi_{r,0}^R(\lambda_{\rm EP}-\tfrac{\epsilon}{2})\rangle,
\end{align} 
where the projection operator $\mathbb{P}_{\lambda_{\rm EP}+\frac{\epsilon}{2}}$ is
\begin{align}
	\mathbb{P}_{\lambda_{\rm EP}+\frac{\epsilon}{2}}=|\psi_{c,0}^R(\lambda_{\rm EP}{+}\tfrac{\epsilon}{2})\rangle\langle \psi_{c,0}^L(\lambda_{\rm EP}{+}\tfrac{\epsilon}{2})| 
	+|\psi_{c,1}^R(\lambda_{\rm EP}{+}\tfrac{\epsilon}{2})\rangle\langle \psi_{c,1}^L(\lambda_{\rm EP}{+}\tfrac{\epsilon}{2})|.
\end{align}
The following identity for the projection operator is derived.
\begin{align}
\lim_{\epsilon\to0}\mathbb{P}_{\lambda_{\rm EP}+\frac{\epsilon}{2}}&=\openone-
\lim_{\epsilon\to0}\sum_{n\geq2}|\psi_n^R(\lambda_{\rm EP}{+}\tfrac{\epsilon}{2})\rangle\langle \psi_n^L(\lambda_{\rm EP}{+}\tfrac{\epsilon}{2})| \nonumber\\
&=\openone-\lim_{\epsilon\to0}\sum_{n\geq2}|\psi_n^R(\lambda_{\rm EP}{-}\tfrac{\epsilon}{2})\rangle\langle \psi_n^L(\lambda_{\rm EP}{-}\tfrac{\epsilon}{2})|\label{eq:projectionIdentity},
\end{align}
where $|\psi_{n\geq2}^{R/L}(\lambda_{\rm EP}{+}\frac{\epsilon}{2})\rangle$ states are excited right and left eigenstates at the parameter $\lambda=\lambda_{\rm EP}+\frac{\epsilon}{2}$. In the derivation of the Eq.\eqref{eq:projectionIdentity}, we assume that there is a finite gap between these excited states $|\psi_{n\geq2}^{R/L}\rangle$ and states $\{ |\psi_{r/c,0}^{R/L}\rangle,\; |\psi_{r/c,1}^{R/L}\rangle\}$ as shown in Fig.\ref{fig:OneHalfFidelity} and there are no exceptional points in the excited states at the same parameter point $\lambda=\lambda_{\rm EP}$.
I.e.,  $\lim\limits_{\epsilon\to0}[|\psi_{n\geq2}^{R/L}(\lambda_{\rm EP}{+}\frac{\epsilon}{2})\rangle - |\psi_{n\geq2}^{R/L}(\lambda_{\rm EP}{-}\frac{\epsilon}{2})\rangle ]=0$.

Using the Eq.~\eqref{eq:projectionIdentity} and the biorthogonality $\langle \psi_{r,0}^L(\lambda_{\rm EP}{-}\frac{\epsilon}{2})|\psi_{n\geq2}^R(\lambda_{\rm EP}{-}\frac{\epsilon}{2})\rangle=0$, we obtain 
\begin{align}\label{eq:one-half}
\lim_{\epsilon\to0}\mathrm{Re}\mathcal{F}(\lambda_\mathrm{EP},\epsilon)=\frac{1}{2}.
\end{align}
\end{proof}

Note that the limiting process in Eq.~\eqref{eq:centerEP} can be more general, for example, taking the parameter $\lambda_\mathrm{EP}{-}\frac{2\epsilon}{3}$ for one side and $\lambda_\mathrm{EP}{+}\frac{\epsilon}{3}$ for another side. The asymmetric limiting procedure is actually employed in the numerical computations in the next two sections.\footnote{For the general asymmetric limiting procedure, one can consider the metricized fidelity $\mathcal{F}(\lambda_{\rm EP},\epsilon;a,b)=
 \langle \psi_{r,0}^L(\lambda_{\rm EP}-a\epsilon)|\psi_{c,0}^R(\lambda_{\rm EP}+b\epsilon)\rangle\langle \psi_{c,0}^L(\lambda_{\rm EP}+b\epsilon)|\psi_{r,0}^R(\lambda_{\rm EP}-a\epsilon)\rangle$ where $a,b>0$. It can be easily checked that the result Eq.~\eqref{eq:one-half} remains the same.}

Before ending this section, two remarks about the properties of one-half are made. First,
for non-interacting systems, the above identity can be applied to each single particle fidelity $f_k$. Especially for two-band models, since a real part of the single particle fidelity between the real-energy state and imaginary-energy state at momentum $k$ is always given by $\frac{1}{2}$, the many-body fidelity is approximated by\footnote{For a general $2\times2$ PT-symmetric matrix, the fidelity is real if the eigenvectors are both PT-unbroken or both PT-broken, as discussed in the end of Sec.~\ref{sec:constraint}. However, the fidelity Eq.~\eqref{eq:centerEP} is complex in this approach.}
\begin{align}\label{eq:one-half-n}
	\lim_{\epsilon\to0}\mathrm{Re}\mathcal{F}(\lambda_{\rm EP},\epsilon)\approx\left(\frac{1}{2}\right)^{n}
\end{align}
where $n$ is number of $k$-points which involve the real-complex energy change.

Second, the Equation \eqref{eq:projectionIdentity} is based on the assumption that the EP is of the usual second-order. For the higher-order EP, e.g. more than two eigenvectors participate in the coalescence, or many coalescences take place at the same EP, the final Eq.~\eqref{eq:one-half} breaks down. In other words, the fidelity at the EP in this approach Eq.~\eqref{eq:centerEP} provides a general property for the second-order EP, and it is useful for verifying whether the EP belongs to second-order or higher-order, by only computing one eigenstate instead of many excited states.
To be precise, if the real part of fidelity is not 1/2, then the higher-order EP is confirmed.
In the following two sections, we apply these general properties to both the non-interacting and interacting systems, respectively.

\section{Non-Hermitian SSH model}\label{sec:SSH}
\begin{figure}[t]
\centering
\includegraphics[width=4.2in]{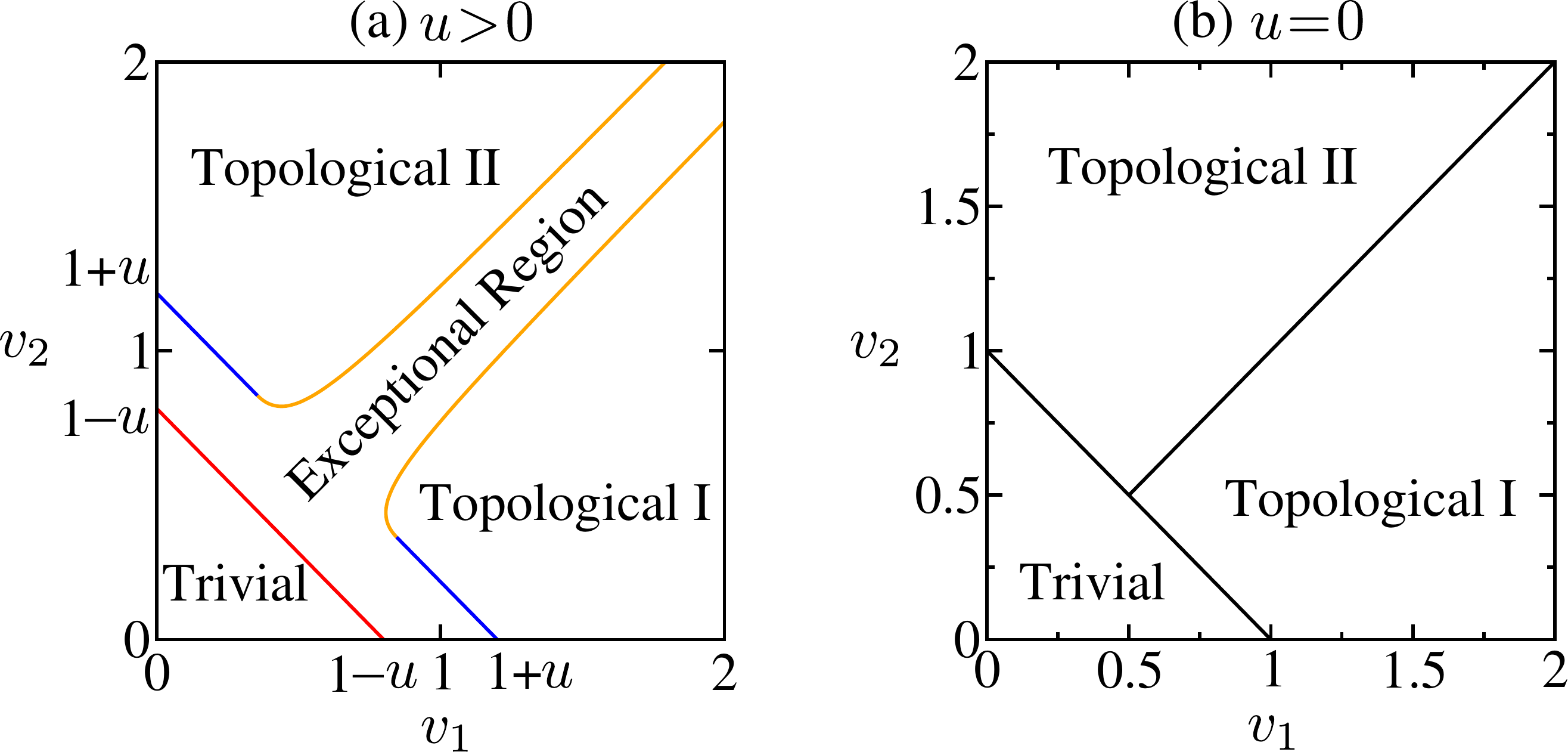}
\caption{Phase diagrams of the SSH model Eq.~\eqref{eq:sshHamiltonian} in the thermodynamic limit, $L\to\infty$. (a) For the presence of the non-Hermiticity, since all the points in the PT-broken phase are EPs, the PT-broken phase is the exceptional region. However, in the finite size system, the number of EPs reduces, as shown in Fig.~\ref{fig:SSHDensityPlot}.
(b) For the absence of non-Hermiticity, the PT-broken phase shrinks into the phase boundaries which separate the topological trivial phase, the topological phase I and II.}
\label{fig:phase}
\end{figure}

The Su-Schrieffer-Heeger (SSH) model is one of the simplest two-band models for studying topological quantum materials. 
It originally describes the electrons moving in the polyacetylene with an alternating hopping term~\cite{SSH1979} and having a wide extension to other topological systems, e.g. the bond-alternating spin-1/2 XXZ chain~\cite{Tzeng2016} and the 3D interacting Weyl semimetals~\cite{Tzeng2020}. 
Due to its simplicity and nontrivial topological properties, it has been treated as a parent model for studying lattice systems with non-Hermiticity~\cite{nSSH_optical_2018,pan2018photonic,Song_2019,nSSH_es1,nSSH}.
Realizing the PT-symmetric non-Hermitian SSH model experimentally is feasible by engineering systems such as the microring laser array~\cite{nSSH_optical_2018} or the waveguide array~\cite{pan2018photonic,Song_2019} with an alternate gain and loss configuration.
The Hamiltonian of the non-Hermitian two-leg SSH ladder is
\begin{align}\label{eq:sshHamiltonian}
    H&=-w\sum_{j=1}^L( c_{j\uparrow}^\dagger c_{j\downarrow} +\mathrm{H.c.})
    -v_1 \sum_{j=1}^L( c_{j\uparrow}^\dagger c_{j+1\downarrow} +\mathrm{H.c.}) \nonumber\\
    &-v_2 \sum_{j=1}^L( c_{j\downarrow}^\dagger c_{j+1\uparrow} +\mathrm{H.c.})  +iu\sum_{j=1}^L(n_{j\uparrow}-n_{j\downarrow}),
\end{align}
where $c_{j\uparrow}$ and $c_{j\downarrow}$ are the fermion annihilation operators at the $j$th unit cell with gain and loss, respectively. $v_1$, $v_2$, and $w$ are the inter- and intra-cell hopping strength, $u$ is the parameter of the non-Hermitian strength, and $n_{j\sigma}=c_{j\sigma}^\dagger c_{j\sigma}$ is the number operator. The model can be regarded as a two-legged ladder, where $\sigma=\uparrow,\downarrow$ is the leg index. Periodic boundary conditions are assumed. 
The Hamiltonian is PT symmetric, where the operator $\mathcal{P}$ acts on the fermion operators as 
$\mathcal{P}c_{j,\uparrow}\mathcal{P}^{-1}=c_{-j,\downarrow}$ and $\mathcal{P}c_{j,\downarrow}\mathcal{P}^{-1}=c_{-j,\uparrow}$,
and $\mathcal{T}$ operator is simply the complex conjugate operator $\mathcal{K}$. 
By employing Fourier transform, $\tilde{c}_{k\sigma}=\frac{1}{\sqrt{L}}\sum_{j=1}^Le^{ikj}c_{j\sigma}$, the Hamiltonian becomes
\begin{align}
    H=\sum_k
    \begin{bmatrix}
        \tilde{c}_{k\uparrow}^\dagger & \tilde{c}_{k\downarrow}^\dagger
    \end{bmatrix}
    \mathcal{H}_k
    \begin{bmatrix}
        \tilde{c}_{k\uparrow}\\
        \tilde{c}_{k\downarrow}
    \end{bmatrix},\quad
    \mathcal{H}_k=\begin{bmatrix}
        iu & \eta \\
        \eta^* & - iu
    \end{bmatrix},
\end{align}
where $k=2\pi m/L$ and $m=0,\dots,L-1$ and $\eta=-w-v_1e^{-ik}-v_2e^{ik}$. 
$\mathcal{H}_k$ is a $2\times2$ PT-symmetric matrix. For each momentum $k$, the $\mathcal{P}\mathcal{T}$ operator is given by $\sigma_x\mathcal{K}$. 
It can be easily checked that $[\mathcal{H}_k,\sigma_x\mathcal{K}]=0$.
After the diagonalization on $\mathcal{H}_k$, the single particle energy band is obtained.
$\varepsilon_{\pm}(k)=\pm\sqrt{\Delta_k}$, where $\Delta_k=v_1^2+v_2^2+w^2+2w(v_1{+}v_2)\cos k + 2v_1v_2\cos{2k}-u^2$.
In the Hermitian limit, i.e. $u=0$, the quantum critical points are located at the parameter where the energy band is gapless.
As the non-Hermitian parameter is turned on, $u>0$, the PT-broken phase appears with non-zero imaginary part of energy.
The phase diagrams in the thermodynamic limit $L\to\infty$ are shown in Fig.~\ref{fig:phase}(a) and (b) for the presence and absence of the non-Hermitian term, respectively.

When the single-particle energies are real, i.e. $\Delta_k>0$, the left and right eigenvectors are
\begin{align}\label{eq:SSHreEeigen}
\ket{L_\pm(k)}&=\frac{1}{\sqrt{2\Delta_k}}
\begin{bmatrix}
    iu\mp\sqrt{\Delta_k}\\
    e^{ik}v_1+e^{-ik}v_2+w
\end{bmatrix}, \notag \\
\ket{R_\pm(k)}&=\frac{1}{\sqrt{2}(\pm iu+\sqrt{\Delta_k})}
\begin{bmatrix}
    -iu\mp\sqrt{\Delta_k}\\
    e^{ik}v_1+e^{-ik}v_2+w
\end{bmatrix}.
\end{align}
When the single-particle energies are complex conjugate pairs, i.e. $\Delta_k<0$, the left and right eigenvectors are
\begin{align}\label{eq:SSHimEeigen}
\ket{L_\pm(k)}&=\mp\sqrt{\frac{u}{-2\Delta_k(u\pm\sqrt{-\Delta_k})}}
\begin{bmatrix}
    iu\pm i\sqrt{-\Delta_k}\\
    e^{ik}v_1+e^{-ik}v_2+w
\end{bmatrix}, \notag \\
\ket{R_\pm(k)}&=\frac{1}{\sqrt{2u(u\pm\sqrt{-\Delta_k})}}
\begin{bmatrix}
    -iu\mp i\sqrt{-\Delta_k}\\
    e^{ik}v_1+e^{-ik}v_2+w
\end{bmatrix}.
\end{align}
They satisfy the bi-orthogonality $\langle L_\pm|R_\pm\rangle{=}1$, $\langle L_\pm|R_\mp\rangle{=}0$ as well as the unit conventional norm of right eigenvectors, $\langle R_\pm|R_\pm\rangle{=}1$.
The half-filled ground state is the product of all the single-particle states with energy $\varepsilon_-(k)$.

In the thermodynamic limit, we can find the exceptional point using the energy band diagram.
Being PT-broken means that $\Delta_k<0$ for some $k$. If we still have $\Delta_k>0$ for some other $k$ (which is the case for this system if we are just across the PT-broken phase transition), then some $k$ must satisfies $\Delta_k=0$. Solving for such $k$ one gets
\begin{align}\label{eq:sol_k}
    \cos k=\frac{-(v_1+v_2)\pm\sqrt{4u^2v_1v_2-(v_1-v_2)^2(4v_1v_2-1)}}{4v_1v_2}.
\end{align}
Here and after, $w=1$ is set to be the unit of the energy scale.
The exceptional point in the thermodynamics limit happens when the solution of $k$ exists but failed to exist after a small change of parameters.
This situation happens in two ways.
The first is $\cos k=-1$, which corresponds to part of the two straight lines $v_1+v_2=1\pm u$.
The region between these lines shrinks to the line $v_1{+}v_2{=}1$ in the Hermitian case. The energy dispersion inside this region is characterized by having one portion of $k$, centered at $\pi$, with imaginary eigenvalues.
The second situation is that the square root in Eq.~\eqref{eq:sol_k} is zero, 
which corresponds to parts of the curve $4u v_1v_2=(v_1-v_2)^2(4v_1v_2-1)$.
The PT-broken phase between these curves shrinks to a line $v_1{=}v_2$ in the Hermitian case, as shown in Fig.~\ref{fig:phase}(b).

\begin{figure}[t]
\centering
    \includegraphics[width=2.8in]{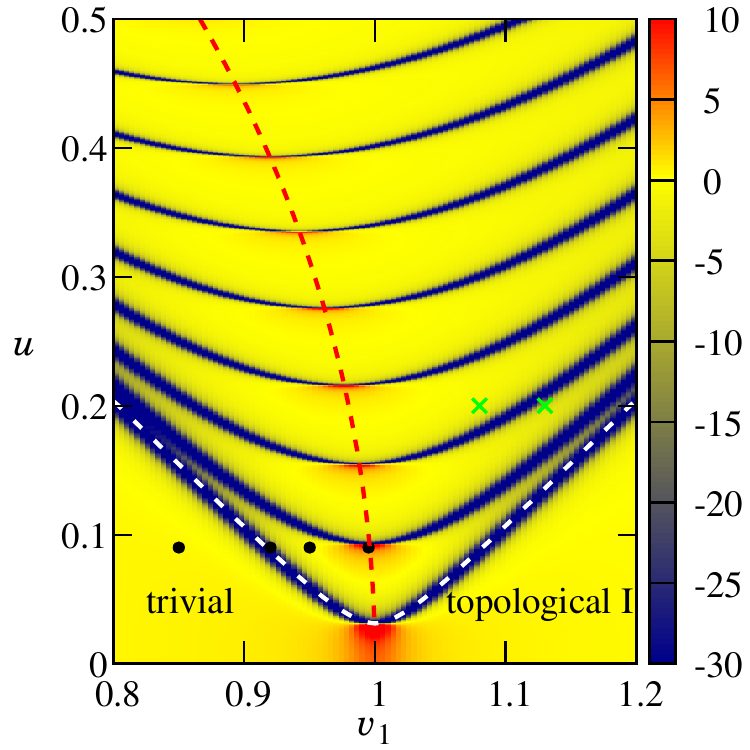} \hspace{0.5cm}
    \includegraphics[width=2.8in]{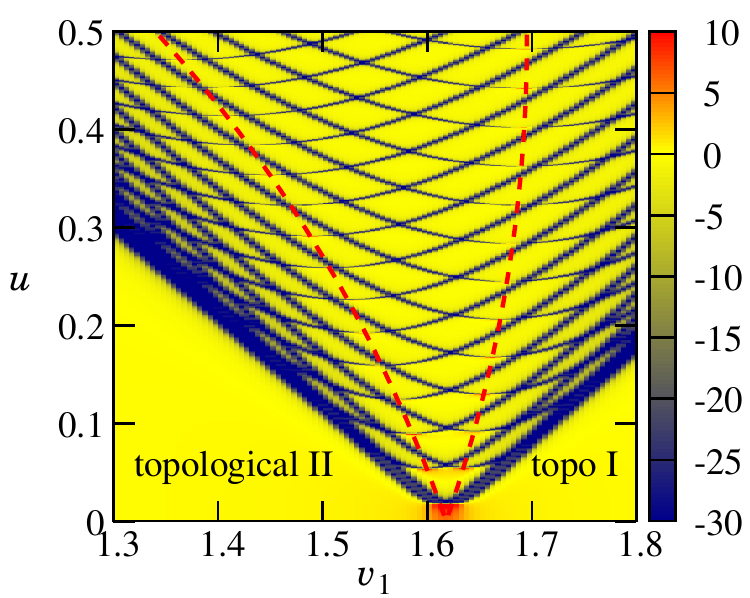}
    \caption{(Left) The fidelity susceptibility density $\mathcal{X}_F/L$ for the finite size SSH model with $v_2=0$ and $L=101$. Below the white dashed line, there are two PT-unbroken phases. Above the white dashed line, it is the PT-broken phase with complex total energy. The EPs are the points inside the blue region with $\mathcal{X}_F/L\to-\infty$. Energy bands $\varepsilon_\pm(k)$ and the single-particle fidelity susceptibilities $\chi_k$ for the four black dots at fixed $u=0.09$ and $v_1=0.85$, $0.92$, $0.95$, and $0.995$ are shown in Fig.~\ref{fig:SSHbandandfsus}. The single-particle fidelity $f_k$ with $\frac{1}{2}$ between the green points $(v_1,u)=(1.08,0.2)$ and $(1.13,0.2)$ is shown in Fig.~\ref{fig:OneHalfSSH}. Finite size scaling for the first red regime is shown in Fig.~\ref{fig:odd-scaling-ssh}.
(Right) $\mathcal{X}_F/L$ for $v_2=\frac{1}{2}(1+\sqrt{5})$ and $L=505$.}\label{fig:SSHDensityPlot}
\end{figure}

\subsection{Fidelity susceptibility}
Using the eigenstates in Eqs.~\eqref{eq:SSHreEeigen}, \eqref{eq:SSHimEeigen} and the definition of the fidelity susceptibility Eq.~\eqref{eq:fsus} or the Eq.~\eqref{eq:bandmomentumfsus}, we obtain the following analytic expression of the fidelity susceptibility for a ground state of the SSH model. 
\begin{subequations}\label{eq:SSHfsus}
	\begin{align}
	\mathcal{X}_F&=\sum_k\chi_k,\\
	\chi_k&=\frac{\sin^2k-u^2+v_2(\cos k-\cos3k)+v_2^2\sin^22k}{4\Delta_k^2}.\label{eq:SSHchik}
\end{align}
\end{subequations}
Here the ground state is defined as a state with all $\varepsilon_-(k)=-\sqrt{\Delta_k}$ states are filled.

To see how the fidelity susceptibility changes across a phase transition point, we consider the phase transition between `Trivial' and `Topological I' shown in Fig.~\ref{fig:phase}. For simplicity, we fix the value of the $v_2$ to zero. The Fig.~\ref{fig:SSHDensityPlot}(left) shows the total susceptibility density $\mathcal{X}_F/L$ as a function of $v_1$ and $u$ with fixed value of $v_2=0$ and the system size $L=101$. The white dashed line in Fig.~\ref{fig:SSHDensityPlot}(left) represents the boundary between the PT-unbroken and PT-broken phases. In the PT-unbroken phase, fidelity susceptibility density shows a peak near the Hermitian critical point $(v_1,u)=(1,0)$, and the peak value is enhanced by the non-Hermitian parameter $u$. More detailed analysis about this peak is given in the appendix. In the PT-broken region, there are many large negative valued-$\mathcal{X}_F/L$ lines inside the dark-blue regime which correspond to collections of EPs which are referred to EP lines. We find that the number of the EP lines increases as the size of the system increases and all the points in the PT-broken phase become EPs in the thermodynamic limit. Besides to the EP lines, it appears some points with a positive valued-$\mathcal{X}_F/L$ (the red regimes) inside the PT-broken region. 

\begin{figure}[t]
\centering
	\includegraphics[width=4.5in]{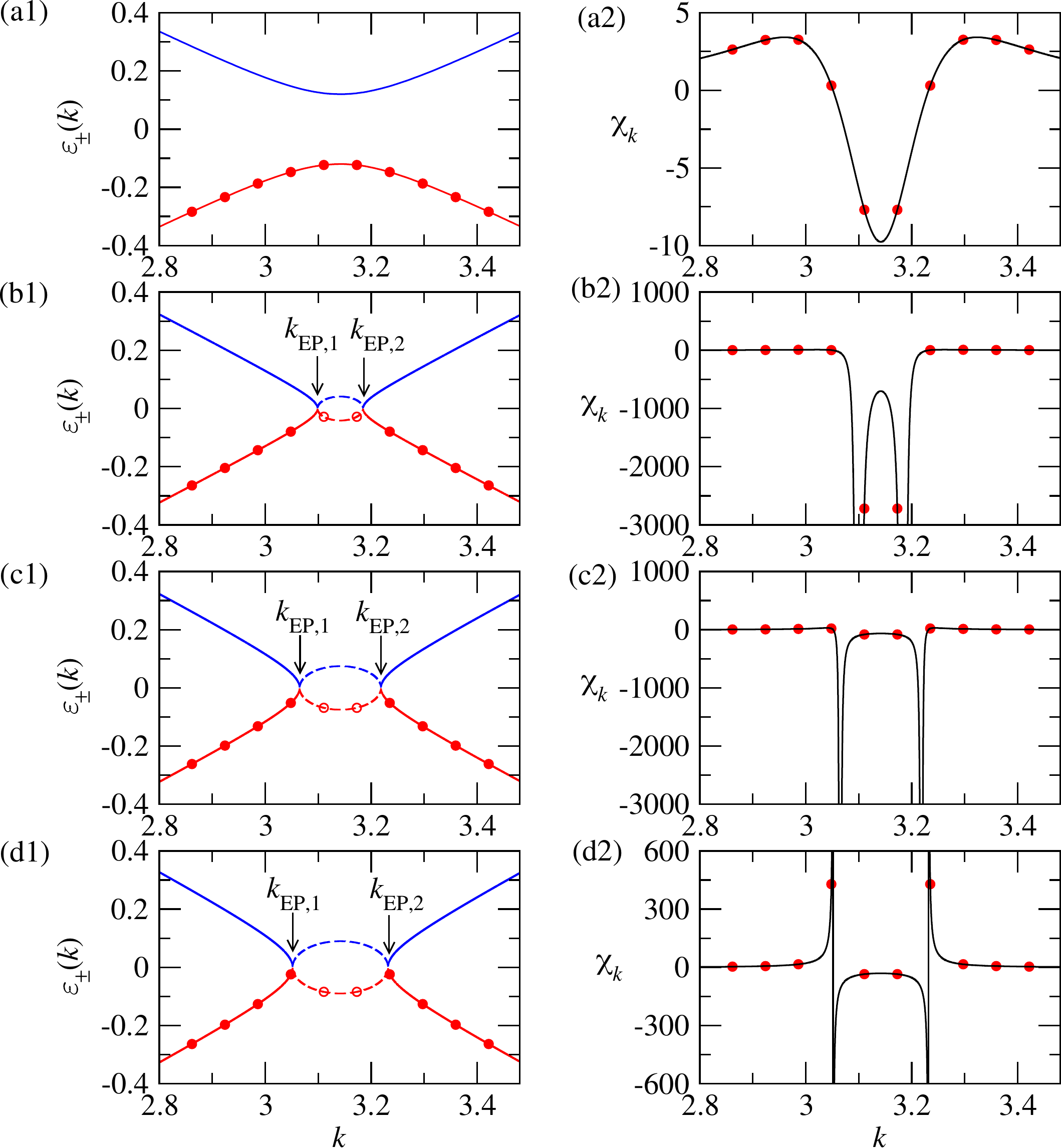}
	\caption{Bands and momentum susceptibilities of four cases; (a) $v_1=0.85$, (b) $v_1=0.92$, (c) $v_1=0.95$ and (d) $v_1=0.995$. The range of the momentum is $k\in (2.8,3.48)$, and $u=0.09$. In the band figures, blue line: $\varepsilon_+(k)$, red line: $\varepsilon_-(k)$, the solid line: non-zero real part of the energies and the dashed line: non-zero imaginary part of the energies. Red dots show values of the energies and susceptibilities at finite momentum values given by $k=\frac{2\pi m}{L}$ where $m=0,\cdots,L-1$ , and $L=101$.}\label{fig:SSHbandandfsus}
\end{figure}

To understand these properties of the $\mathcal{X}_F/L$, we consider band structures $\varepsilon_{\pm}(k)$ and the single particle fidelity susceptibility $\chi_k$ Eq.~\eqref{eq:SSHfsus} at the four points shown in Fig.~\ref{fig:SSHDensityPlot}(left). The resulting bands and single particle fidelity susceptibilities are shown in Fig.~\ref{fig:SSHbandandfsus}. 

At the first point $(u,v_1)=(0.09,0.85)$, it belongs to the PT-unbroken phase, and all momentum points have real energies, as shown in Fig.\ref{fig:SSHbandandfsus}~(a1). Therefore ground state energy is real. Since this point is far from both the quantum critical point and EPs, magnitudes of $\chi_k$ are small over all $k$-points. As the parameter $v_1$ increases, the gap becomes closer and they meet at $k=\pi$ and the two $k_{\rm EP}$'s appear as shown in Fig.~\ref{fig:SSHbandandfsus}. In the thermodynamic limit, this gap closing point corresponds to the boundary between PT-unbroken phase and the PT-broken phase. However, since we are considering the finite system with $L=101$, $k$-points shown in Fig.~\ref{fig:SSHbandandfsus} with solid dots and empty dots are considered only to contribute in the fidelity susceptibility. As a result, the boundary between the PT-unbroken and the PT-broken phase depends on the number of unit cells. For example, for the odd number of unit cells here ($L=101$), two momentum points around $\pi$, i.e. $k=\pi-\frac{\pi}{L}$, and $k=\pi+\frac{\pi}{L}$, (the empty dots in Fig.~\ref{fig:SSHbandandfsus}) are filled by the particles. If the distance between these two momentum points is larger than the distance between two exceptional momentum points ($=k_{\rm EP,2}-k_{\rm EP,1}$), then the ground state energy is given by real value numerically. Therefore we identify this state as a PT-unbroken state. However, if the system size $L$ larger than the $L_0=\frac{2\pi}{|k_{\rm EP,2}-k_{\rm EP,1}|}$, then the ground state energy is complex, and this state is identified as a PT-broken state.

In the cases of the other three parameter points, $(u,v_1)=(0.09,0.92)$, $(0.09,0.95)$, and $(0.09,0.995)$, they have similar band structures, i.e. two of the $k$-points with imaginary energies, as shown in Fig.~\ref{fig:SSHbandandfsus}(b1), (c1) and (d1). Therefore they belong to the PT-broken phase. However, the values $\mathcal{X}_F/L$ are quite different. At $(u,v_1)=(0.09,0.92)$, there are two $k$-points with imaginary energies closing to the $k_\mathrm{EP}$'s and contributing to large negative susceptibility, as shown in Fig.~\ref{fig:SSHbandandfsus}(b2). While for $(u,v_1)=(0.09,0.95)$, these two $k$-points are far from the $k_\mathrm{EP}$'s, and have small contributions to the susceptibility, as shown in Fig.~\ref{fig:SSHbandandfsus}(c2). 
At the last point $(u,v_1)=(0.09,0.995)$, the $\mathcal{X}_F/L$ shows a positive value inside the PT-broken phase which corresponds to the second red regime in Fig.~\ref{fig:SSHDensityPlot}(left). It shows a large positive single-particle fidelity susceptibility when $k$ approaches to $k_\mathrm{EP}$'s from the real energy side, as shown in Fig.~\ref{fig:SSHbandandfsus}(d1) and (d2). Note that the $\chi_k$ always negative for the imaginary single-particle energy. In the following, we argue these positive peaks in the PT-broken phase are related to the Hermitian quantum critical point.

From Eq.~\eqref{eq:SSHchik} it is observed that, the sign of $\chi_k$ is determined by the numerator because the denominator $4\Delta_k^2\geq0$. As $k$ varies, the function $\chi_k$ changes the sign when the numerator is zero, i.e. $\sin^2k-u^2+v_2(\cos k-\cos3k)+v_2^2\sin^2{2k}=0$. Together with the fact that the divergence takes place when the denominator is zero, i.e. $\Delta_k=0$, the condition of the positive divergence of total fidelity susceptibility in Fig.~\ref{fig:SSHDensityPlot} is both the numerator and denominator in Eq.~\eqref{eq:SSHchik} are zero. The solution of these equations for the $v_2=0$ is shown in the Fig.~\ref{fig:SSHDensityPlot}(left) as the red dashed curve, $u=\sqrt{1-v_1^2}$. For $v_2=(1+\sqrt{5})/2$, the red dashed curve in Fig.~\ref{fig:SSHDensityPlot}(right), the parametric formula of the solution is $v_1=-\cos{k}-\frac{1}{2}(1+\sqrt{5})\cos2k$, $u=\sqrt{[4+\sqrt{5}+2(1+\sqrt{5})\cos k +(3+\sqrt{5})\cos2k]\sin^2{k}}$, for $k\in[0,2\pi]$. For both the cases $v_2=0$ and $v_2\neq0$, the red dashed curves pass though the Hermitian quantum critical points, therefore we speculate the positive divergence of the fidelity susceptibility in the PT-broken phase is the extension from the Hermitian quantum critical points.

\subsection{Fidelity of 1/2 at EP}
\begin{figure}[t]
\centering
\includegraphics[width=4in]{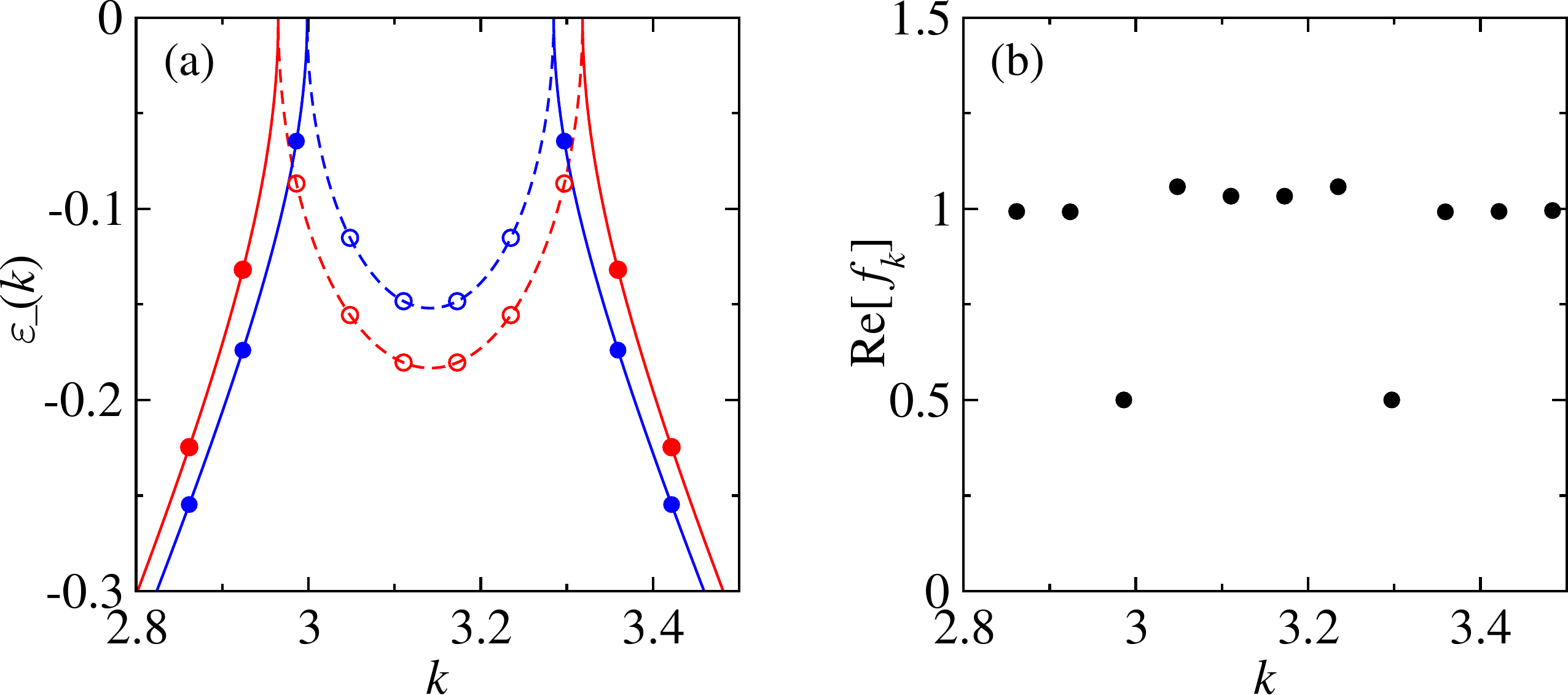}
\caption{(a) The single-particle energy $\varepsilon_-(k)$ in the SSH model for fixed $u=0.2$, $v_2=0$, and (red) $v_1=1.08$ and (blue) $v_1=1.13$, corresponding to the green points in Fig.~\ref{fig:SSHDensityPlot}. The solid/empty dots represent the finite size $k$ points with real/imaginary energies for $L=101$. The EP is between $v_1=1.08$ and $v_1=1.13$. (b) The single-particle fidelity $f_k$ between these two states. At the $k$-points where $\varepsilon_-(k)$ change from real to imaginary, the fidelities are complex and the real part is $\frac{1}{2}$.}\label{fig:OneHalfSSH}
\end{figure}

We demonstrate an example of Eq.~\eqref{eq:one-half-n} in the SSH model by the single-particle picture. The single-particle fidelity $f_k$ between the green points in Fig.~\ref{fig:SSHDensityPlot} is calculated directly by the product of the single-particle eigenvectors. The parameters are $(v_1,u)=(1.08,0.2)$ and $(1.13,0.2)$, and the system size is $L=101$. As shown in Fig.~\ref{fig:OneHalfSSH}(a), the single-particle energies $\varepsilon_-(k)$ for these two parameter settings are compared. There are two $k$-points that encounter the real-imaginary energy changes. One can imagine that as the parameter goes from $v_1=1.08$ to $1.13$, the $k_{\rm EP}$'s pass through these two $k$-points. The single-particle fidelity for these two $k$-points becomes the product of PT-unbroken state Eq.~\eqref{eq:SSHreEeigen} and PT-broken state Eq.~\eqref{eq:SSHimEeigen}, and the real part $\mathrm{Re}[f_k]=\frac{1}{2}$, as shown in Fig.~\ref{fig:OneHalfSSH}(b). In this case, the total fidelity is approximated by $\mathrm{Re}\mathcal{F}\approx(\frac{1}{2})^2$, verifying that the number of second-order $k_{\rm EP}$ is $2$.

The analysis for the transition between `Topological~I' to `Topological~II' is similar to $v_2=0$.
In this case, changes of the band structures are more complicated and it turns out that portions of the discrete momentum points belong to momentum regions which have imaginary energies changes in a chaotic manner depending on the size $L$. For simplifying the analysis, we consider that the two portions of $k$ with imaginary energies are centered at $k=4\pi/5$ and $k=6\pi/5$ for small $u$. This corresponds to $v_2=\frac{1}{2}(1+\sqrt{5})$. With this choice, if $L=10n+5$, $n=0,1,2,\ldots$, the behavior is similar to the phase transition between `Trivial' to `Topological I' phases. In other words, there is an $L_0$, and the finite size ground state is still PT-unbroken when $L<L_0$. 
Although there are more subtleties in analyzing this case, there is no qualitative difference compared to the `Trivial' to `Topological~I' transition case. Therefore, we only present $\mathcal{X}_F/L$ for $v_2=\frac{1}{2}(1+\sqrt 5)$ in Fig.~\ref{fig:SSHDensityPlot}(right) without showing detailed analysis.

We have demonstrated the fidelity susceptibility in the non-interacting non-Hermitian system with the understanding via the single-particle picture. For interacting many-body systems, where the single-particle picture does not apply, we employ numerical methods on finite-size systems in the next section.

\section{Non-Hermitain spin-1/2 XXZ chain}\label{sec:XXZ}
The Hermitian XXZ model is a well-known one-dimensional spin model that has received significant attention in the field of condensed matter physics~\cite{Mikeska2004}. It is a generalization of the quantum Heisenberg model, incorporating an anisotropy term that breaks the SU(2) spin rotational symmetry. Nevertheless, the Hamiltonian still has U(1) continuous symmetry, and the magnetization $M=\sum_j\sigma_j^z$ is conserved, i.e. $[H,M]=0$. In the Hermitian limit, the XXZ model has a rich phase diagram that has been extensively investigated, including the case with bond alternation~\cite{Tzeng2016}, which can be connected to the SSH model with density-density nearest-neighbor interaction.
In this section, we consider the non-Hermitian spin-1/2 XXZ model by adding the staggered imaginary magnetic field in $z$-direction. The Hamiltonian is given by
\begin{align}\label{eq:xxz}
  H = \sum_{j=1}^L \left(\sigma_j^x\sigma_{j+1}^x + \sigma_j^y\sigma_{j+1}^y + J_z \sigma_j^z\sigma_{j+1}^z\right)
    + i\gamma \sum_{j=1}^{L/2} \left( \sigma_{2j-1}^z - \sigma_{2j}^z \right),
\end{align}
where $\vec{\sigma}_j$ are the Pauli matrices at the site $j$.
The model Eq.~\eqref{eq:xxz} with $J_z=0$ is equivalent to the non-interacting SSH model Eq.~\eqref{eq:sshHamiltonian} with $v_1=w=1$, $v_2=0$, and $u=\gamma$, by the Jordan-Wigner transformation. 
Note that a related variant of the non-Hermitian XXZ model has been proposed recently, which may be realized in the ultra-cold atom experiments~\cite{Ashida2017}.

\begin{figure}[t]
\center
\includegraphics[width=4in]{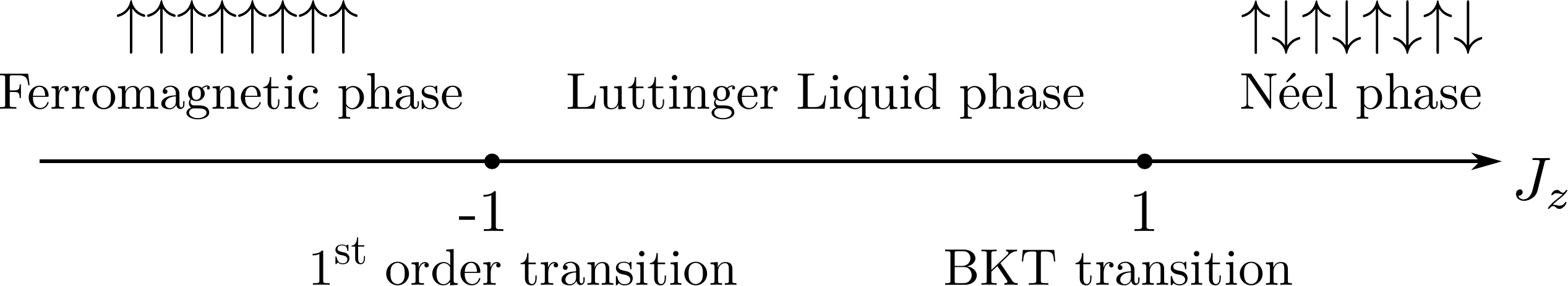}
\caption{The ground state phase diagram of the 1D spin-1/2 XXZ model in the Hermitian limit, i.e. $\gamma=0$.}\label{fig:XXZphase}
\end{figure}

It is worth providing a brief introduction to the phase diagram of the Hermitian ($\gamma=0$) XXZ model, as shown in Fig.~\ref{fig:XXZphase}, and the significance of studying fidelity susceptibility on this model. The Luttinger liquid phase, where the ground state is gapless, separates the ferromagnetic and N\'eel phases with $-1<J_z\leq1$. At $J_z=-1$, the system undergoes a first-order transition from the Luttinger liquid phase to the ferromagnetic phase, characterized by a sudden change in the magnetization. The ferromagnetic ground state is $\mid\cdots\uparrow\uparrow\uparrow\uparrow\cdots\rangle$ or $\mid\cdots\downarrow\downarrow\downarrow\downarrow\cdots\rangle$. On the other hand, at $J_z=1$, the system undergoes a Berezinskii-Kosterlitz-Thouless (BKT) quantum phase transition, where the ground state energy and the first excited state energy in the finite size have different parity quantum numbers and become degenerate in the thermodynamic limit~\cite{Tzeng2012}. The BKT transition is a topological phase transition without breaking continuous symmetry, and it is an infinite-order phase transition, which means that the $n$-th derivative of the ground state energy diverges only when $n$ goes to infinity. 
When $J_z>1$ is finite, the ground state is not exactly the N\'eel state $\mid\cdots\uparrow\downarrow\uparrow\downarrow\cdots\rangle$ or $\mid\cdots\downarrow\uparrow\downarrow\uparrow\cdots\rangle$ due to the presence of spin-spin exchange coupling in the $x$- and $y$-directions. These couplings cause the ground state to deviate from the N\'eel state, and therefore, the N\'eel state is not an eigenstate of the Hamiltonian. The ground state approximate to the N\'eel state when $J_z\gg1$.
Note that even in the Hermitian case, the N\'eel state is a PT-broken state with real energy. Therefore, it is expected that the ground state energy for $J_z\gg1$ becomes complex when the non-Hermitian term $\gamma>0$ is turned on.

\begin{figure}[t]
\centering
\includegraphics[width=4.3in]{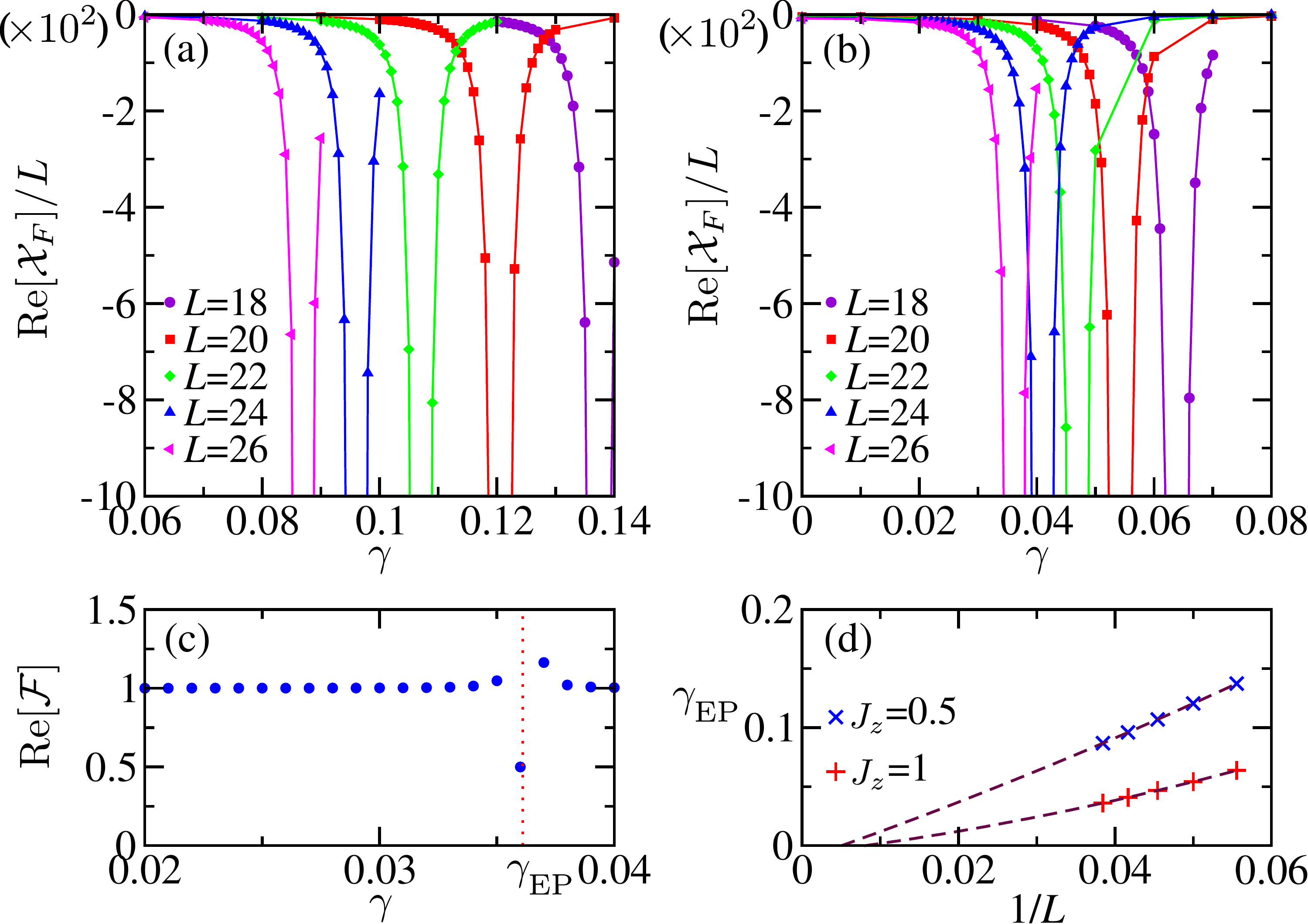}
\caption{The real part of the fidelity susceptibility density $\mathrm{Re}[\mathcal{X}_F]/L$ for the finite size non-Hermitian XXZ model between $\gamma$ and $\gamma+\epsilon$ for fixed (a) $J_z=0.5$ and (b) $J_z=1$. We take $\epsilon=10^{-3}$. The EPs are the points with $\mathrm{Re}[\mathcal{X}_F]/L\to-\infty$.
(c)~The real part of fidelity for $J_z=1$ and $L=26$. The dotted line represents the finite-size EP $\gamma_{\rm EP}$ which separates the PT-unbroken and PT-broken phases. The EPs are identified as the second-order EPs since $\mathrm{Re}[\mathcal{F}]\approx0.5$ when $\gamma<\gamma_{\rm EP}<\gamma+\epsilon$.
(d) The extrapolation of the EPs suggests that for all $J_z\geq0$, a finite non-Hermitian strength $\gamma>0$ brings the system into the PT-broken phase in the thermodynamic limit.}\label{fig:one-half-xxz}
\end{figure}

The fidelity susceptibility has been developed as a nice probe to locate the second-order transitions. For higher order transitions, it is shown that the fidelity susceptibility still works for censoring the third-order transitions. However, it fails for the fifth-order Gaussian quantum phase transitions~\cite{Tzeng2008b}. Surprisingly, Yang~\cite{MFYang2007} and Fj{\ae}restad~\cite{Fj_restad_2008} show in the thermodynamic limit with different analytical approaches, respectively, that the fidelity susceptibility density in the Luttinger liquid phase of the Hermitian XXZ model is $[8(\pi-\cos^{-1}J_z)^2(1-J_z)^2]^{-1}$. This indicates the positive divergence of the fidelity susceptibility takes place at both the first-order and the BKT quantum critical points $J_z=\pm1$. However, numerical investigation with the fidelity susceptibility on the BKT quantum critical point is particularly difficult because of the slow divergence behavior with system size~\cite{fsus_BKT_2010, fsus_BKT_2011, fsus_BKT_2015, zhang2021fidelity}. In the previous section, we have observed that the positive peak of $\mathrm{Re}[\mathcal{X}_F]/L$ in the PT-broken phase may be related to the Hermitian quantum phase transitions. 
In the non-Hermitian XXZ model with a sufficiently large $\gamma>0$, we anticipate observing positive peaks of $\mathrm{Re}[\mathcal{X}_F]/L$ in the PT-broken phase. These peaks are expected to stem from one of the Hermitian quantum critical points at $J_z=\pm1$.

We now direct our focus toward the non-Hermitian XXZ model, Eq.~\eqref{eq:xxz}. We initiate our study from the Luttinger liquid phase and introduce non-Hermiticity by traversing along $\gamma$-direction with fixed $J_z=0.5$ and $J_z=1$ to identify the EP using the fidelity susceptibility as a concrete application example of the properties discussed in Sec.~\ref{sec:fsus}. To compute the fidelity, one left/right eigenvector is required, so we employ the Lanczos iteration method in the subspace of $M=0$ to obtain the ground state. The Lanczos iteration method is a well-established numerical algorithm for finding the eigenvalues and eigenvectors of large sparse matrices. In our case, we use the Lanczos method for the complex symmetric matrix, which iteratively constructs a complex orthogonal basis for the Krylov subspace, spanned by the vectors obtained by repeatedly applying the matrix to an initial vector. The Lanczos method is especially useful when the matrix is too large to be stored in memory or when only a few eigenvalues and eigenvectors are needed. The largest system size in this section is upto $L=30$, and the matrix dimension in the $M=0$ subspace is about $1.5\times10^8$.

\begin{figure}[t]
\includegraphics[width=6.5in]{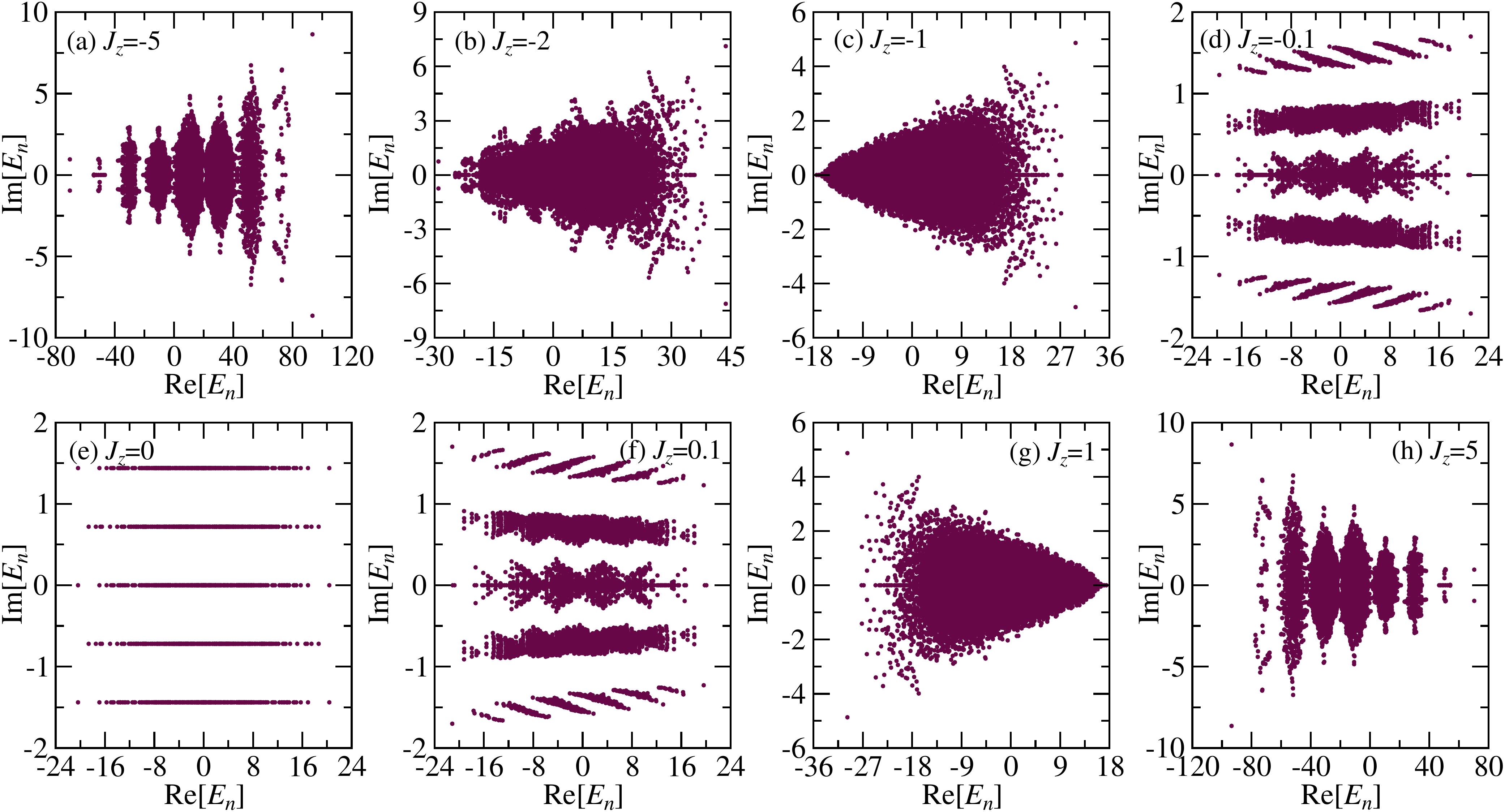}
\caption{The full energy spectra in the subspace $M=0$ of the non-Hermitian XXZ spin-1/2 chain with $\gamma=0.5$ and different $J_z$. The system size is $L=18$. These complex energy spectra indicate that $\gamma=0.5$ is sufficient to drive the system into the PT-broken phase.}\label{fig:XXZfullED18}
\end{figure}

The fidelity susceptibility densities for $J_z=0.5$ and $J_z=1$ as functions of $\gamma$ for the finite size systems are shown in Fig.~\ref{fig:one-half-xxz}(a) and (b), respectively. For small $\gamma\ll1$, both the ground state energy and $\mathcal{X}_F/L$ are real, and the state is PT-unbroken until $\gamma$ crosses the EP $\gamma_{\rm EP}$. When $\gamma>\gamma_{\rm EP}$, it is in the PT-broken phase, and both the ground state energy and $\mathcal{X}_F/L$ have nonzero imaginary parts, although only the real part is shown. From both the PT-unbroken and PT-broken sides approach to $\gamma_{\rm EP}$, a clear divergence of $\mathrm{Re}[\mathcal{X}_F]/L\to-\infty$ is found. During the parameter scan of $\gamma$, it is observed, as shown in Fig.~\ref{fig:one-half-xxz}(c), that the real part of the fidelity is approximately equal to $\mathrm{Re}[\mathcal{F}]\approx\frac{1}{2}$ when $\gamma<\gamma_{\rm EP}<\gamma+\epsilon$, where $\epsilon=10^{-3}$. This property of the fidelity, obtained by considering only the ground state instead of many excited states, suggests that the EP is of second order.
However, the location of the EP shifts as the system size changes. In Fig.~\ref{fig:one-half-xxz}(d), the extrapolation is performed. For both $J_z=0.5$ and $J_z=1$, since the extrapolated EPs are slightly less than zero, we believe that in the thermodynamic limit $L\to\infty$, for $J_z\geq0$ and $\gamma>0$, the ground state is PT-broken. Note that for $J_z=0$ the result is consistent with the SSH model in the previous section. 
Next, we focus on analyzing the behavior of the system in the PT-broken phase along the $J_z$-direction.

\begin{figure}[t]
\centering
\includegraphics[width=5.7in]{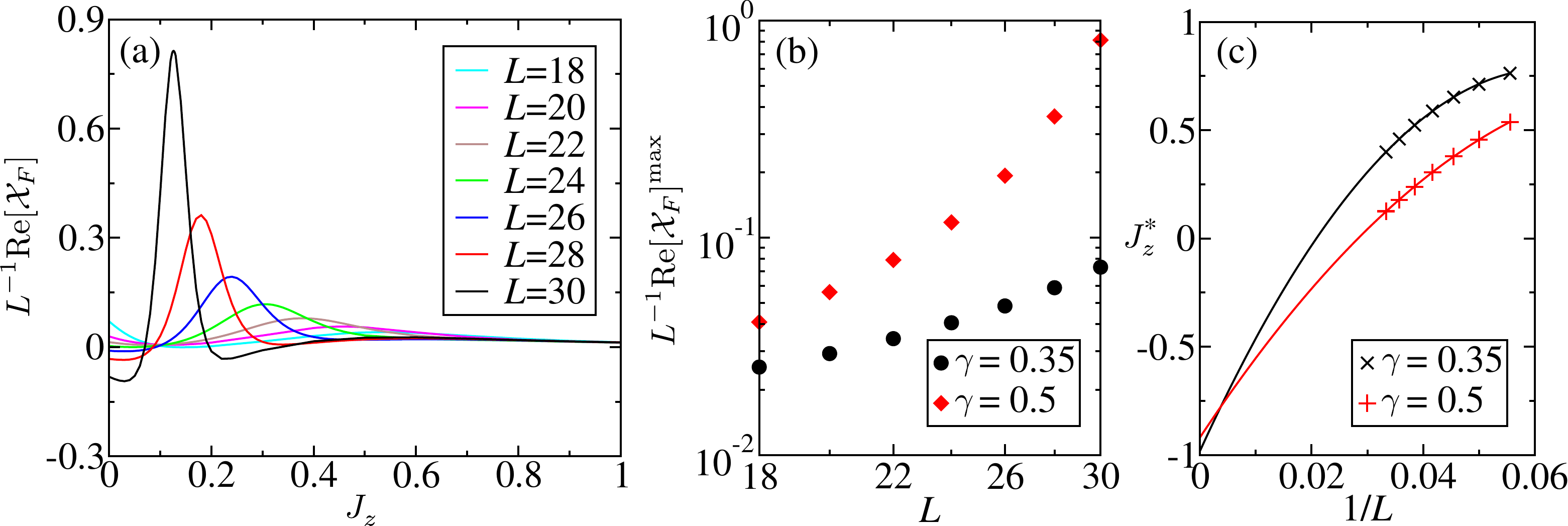}
\caption{(a) The real part of fidelity susceptibility density $\mathrm{Re}[\mathcal{X}_F]/L$ between $J_z$ and $J_z+\epsilon$ for the XXZ model with fixed $\gamma=0.5$. We take $\epsilon=10^{-3}$. In the PT-broken phase, the positive peaks are clearly shown. (b) Log-log scale plot of the maximum of the peak values versus system sizes shows that the peak grows faster than a power-law as size increases. (c) Extrapolation with polynomial fitting suggests the positive peak comes from the first-order transition at $J_z=-1$ of the Hermitian XXZ chain.}\label{fig:1st-order}
\end{figure}

The energy spectra of the non-Hermitian XXZ spin-1/2 chain with a system size of $L=18$ and fixed $\gamma=0.5$ are investigated by using full diagonalization in the subspace $M=0$ for different values of $J_z$. As shown in Fig.~\ref{fig:XXZfullED18}, the complex energy spectra reveal that the system has been driven into the PT-broken phase by the $\gamma=0.5$. We have observed variations in the energy spectra with varying $J_z$ values. Specifically, we note differences in the quantization of energy levels, with some levels being imaginary energy quantized and others being real energy quantized. These observations suggest a significant change in the behavior of the system, highlighting the need for further investigation. We have verified that a smaller value of $\gamma=0.35$ is also sufficient to stay in the PT-broken phase. In the following, we present the fidelity susceptibility analysis in the PT-broken phase along the $J_z$-direction with fixed $\gamma=0.5$ and $0.35$, which forms the primary focus of this paper.

Figure~\ref{fig:1st-order} presents the results of the real part of fidelity susceptibility density $\mathrm{Re}[\mathcal{X}_F]/L$ for the non-Hermitian XXZ model with fixed $\gamma=0.35$ and $\gamma=0.5$. In Fig.~\ref{fig:1st-order}(a), the positive peaks of $\mathrm{Re}[\mathcal{X}_F]/L$ for different system sizes up to $L=30$ are clearly shown. We label the maximum values $\mathrm{Re}[\mathcal{X}_F]^\mathrm{max}$ and their positions $J_z^*$ for each size and perform scaling analysis. The positive peaks tend to diverge faster than a power-law as system size increases, which is shown in the logarithmic scale plot in Fig.~\ref{fig:1st-order}(b). Finally, Fig.~\ref{fig:1st-order}(c) shows the results of extrapolation with polynomial fitting for the position, which yields $J_z\approx -1$ for both $\gamma=0.35$ and $\gamma=0.5$, suggesting that the positive peaks in the PT-broken phase may originate from the first-order transition point rather than the BKT point.
Given the limited scope of exact diagonalization in the case of small finite sizes, the positive peaks that arise from the BKT quantum critical point may not be discernible. As such, it would be of great interest to explore alternative numerical methodologies with more sophisticated techniques to further elucidate the nature of these peaks.

\section{Discussion and Conclusion}\label{sec:conclusion}
Requiring that the fidelity be real and positive for a non-Hermitian quantum system is overly restrictive.
Since the more physical condition of the PT-symmetry is reasonably considered in the complex generalization of quantum mechanics~\cite{Bender1998,Bender2002}, 
the proper definition of fidelity with the less restrictive conditions in the biorthogonal basis is now taking the PT-symmetry into account.
We have derived the general properties of the fidelity Eq.~\eqref{eq:fLR} and its susceptibility Eq.~\eqref{eq:fsus} by the perturbation theory with additional constrains of PT-symmetry. 
We prove that the fidelity susceptibility is always real if the states are PT-unbroken. 
For the PT-broken states, the real part of the fidelity susceptibility deserves attention, because it is equivalent to considering with both the PT-partner states. 
Especially, when the parameter approaches to the EP from the PT-broken state, the conjecture of the negative divergence Eq.~\eqref{eq:negative_infty} is supported. 
We prove another important fundamental property of the second-order EP that the real part of the fidelity between PT-unbroken and PT-broken states is $\mathrm{Re}\mathcal{F}=\frac{1}{2}$.
We provide concrete examples of the above properties in both non-interacting and interacting many-body systems. Moreover, it is found that the positive divergence of fidelity susceptibility density at the Hermitian quantum critical point has some extensions into the non-Hermitian PT-broken phase.

The fidelity is especially beneficial for numerical investigation, as many numerical methods, e.g. the Lanczos exact diagonalization method or the tensor network methods~\cite{Tzeng2012,TensorNetwork}, for obtaining only the ground state are easier than obtaining many excited states. Without the information about the excited states, the fidelity and fidelity susceptibility serve as theoretical tools for locating the EP, and verifying whether it is a higher-order EP.

Note that very recently the negative divergence property Eq.~\eqref{eq:negative_infty} has been applied to the non-Hermitian quantum many-body scar~\cite{scar}. On the other hand, fidelity susceptibility without showing positive or negative peaks in the Affleck-Kennedy-Lieb-Tasaki (AKLT) model provide evidence of Symmetry-Protected-Topological phase does not have either quantum critical point nor EP within small non-Hermitian parameter strength~\cite{tu2021renyi}. Therefore, it is believed that the fidelity Eq.~\eqref{eq:fLR} and its susceptibility Eq.~\eqref{eq:fsus} are also useful for other non-Hermitian systems.

Before closing the discussion, we emphasize one unique feature in the non-interacting cases. 
In the thermodynamic limit, one of the single-particle states will be at EP in the PT-broken phase and the PT-broken phase becomes
the exceptional region. I.e., the fidelity susceptibility will be infinite negative in the entire PT-broken phase in Fig.~\ref{fig:SSHDensityPlot}.
The red dashed lines in Fig.~\ref{fig:SSHDensityPlot} suggest a phase boundary inside the exceptional region and how to characterize phases in the exceptional region is an interesting question for future investigation.\footnote{In App.~\ref{app:Boundary}, we find the red dashed line is not the phase boundary between the topological and trivial phases. Here we define the topological and trivial phases from the existence of the boundary modes for open boundary condition. The complex Berry phase~\cite{nSSH, Liang2013} also cannot detect the red dashed line in the non-Hermitian SSH model. It might suggest the red dashed line is the boundary between two distinct EP states and these EP states cannot be diagnosed from the boundary modes or the complex Berry phase.}

\begin{acknowledgments}
Yi-Ting Tu and Iksu Jang contributed equally to this work. We are grateful to Jhih-Shih You for many useful discussions. PYC is supported by National Science and Technology Council (NSTC) in Taiwan, under grant No. NSTC 112-2636-M-007-007. Y.C.T. and P.Y.C. acknowledge the support from the National Center for Theoretical Sciences (NCTS) in Taiwan.
\end{acknowledgments}


\begin{thebibliography}{10}

\bibitem{You2007}
W.-L. You, Y.-W. Li, and S.-J. Gu.
\newblock ``Fidelity, dynamic structure factor, and susceptibility in critical phenomena''.
\newblock \href{https://doi.org/10.1103/PhysRevE.76.022101}{Phys. Rev. E {\bf 76}, 022101}~(2007).

\bibitem{fidelity_review}
S.-J. Gu.
\newblock ``Fidelity approach to quantum phase transitions''.
\newblock \href{https://doi.org/10.1142/S0217979210056335}{Int. J. Mod. Phys. B {\bf 24}, 4371--4458}~(2010).

\bibitem{MFYang2007}
M.-F. Yang.
\newblock ``Ground-state fidelity in one-dimensional gapless models''.
\newblock \href{https://doi.org/10.1103/PhysRevB.76.180403}{Phys. Rev. B {\bf 76}, 180403(R)}~(2007).

\bibitem{Fj_restad_2008}
John~Ove Fj{\ae}restad.
\newblock ``Ground state fidelity of {L}uttinger liquids: a wavefunctional approach''.
\newblock \href{https://doi.org/10.1088/1742-5468/2008/07/p07011}{Journal of Statistical Mechanics: Theory and Experiment {\bf 2008}, P07011}~(2008).

\bibitem{Tzeng2008a}
Y.-C. Tzeng and M.-F. Yang.
\newblock ``Scaling properties of fidelity in the spin-1 anisotropic model''.
\newblock \href{https://doi.org/10.1103/PhysRevA.77.012311}{Phys. Rev. A {\bf 77}, 012311}~(2008).

\bibitem{Tzeng2008b}
Y.-C. Tzeng, H.-H. Hung, Y.-C. Chen, and M.-F. Yang.
\newblock ``Fidelity approach to {G}aussian transitions''.
\newblock \href{https://doi.org/10.1103/PhysRevA.77.062321}{Phys. Rev. A {\bf 77}, 062321}~(2008).

\bibitem{Ren_2015}
J. Ren, G.-H. Liu, and W.-L. You.
\newblock ``Entanglement entropy and fidelity susceptibility in the one-dimensional spin-1 {XXZ} chains with alternating single-site anisotropy''.
\newblock \href{https://doi.org/10.1088/0953-8984/27/10/105602}{J. Phys.: Cond. Mat. {\bf 27}, 105602}~(2015).

\bibitem{fsus_BKT_2010}
B. Wang, M. Feng, and Z.-Q. Chen.
\newblock ``{B}erezinskii-{K}osterlitz-{T}houless transition uncovered by the fidelity susceptibility in the \textit{XXZ} model''.
\newblock \href{https://doi.org/10.1103/PhysRevA.81.064301}{Phys. Rev. A {\bf 81}, 064301}~(2010).

\bibitem{fsus_BKT_2011}
H.-L. Wang, J.-H. Zhao, B.~Li, and H.-Q. Zhou.
\newblock ``{K}osterlitz-{T}houless phase transition and ground state fidelity: a novel perspective from matrix product states''.
\newblock \href{https://doi.org/10.1088/1742-5468/2011/10/l10001}{J. Stat. Mech.: Theo. Exp. {\bf 2011}, L10001}~(2011).

\bibitem{fsus_BKT_2015}
G.~Sun, A.~K. Kolezhuk, and T.~Vekua.
\newblock ``Fidelity at {B}erezinskii-{K}osterlitz-{T}houless quantum phase transitions''.
\newblock \href{https://doi.org/10.1103/PhysRevB.91.014418}{Phys. Rev. B {\bf 91}, 014418}~(2015).

\bibitem{zhang2021fidelity}
J. Zhang.
\newblock ``Fidelity and entanglement entropy for infinite-order phase transitions''.
\newblock \href{https://doi.org/10.1103/PhysRevB.104.205112}{Phys. Rev. B {\bf 104}, 205112}~(2021).

\bibitem{review-Ashida2020}
Y. Ashida, Z. Gong, and M. Ueda.
\newblock ``Non-{H}ermitian physics''.
\newblock \href{https://doi.org/10.1080/00018732.2021.1876991}{Advances in Physics {\bf 69}, 249--435}~(2020).

\bibitem{Ganainy2018}
Ramy El-Ganainy, Konstantinos~G. Makris, Mercedeh Khajavikhan, Ziad~H. Musslimani, Stefan Rotter, and Demetrios~N. Christodoulides.
\newblock ``Non-{H}ermitian physics and {PT} symmetry''.
\newblock \href{https://doi.org/10.1038/nphys4323}{Nature Physics {\bf 14}, 11--19}~(2018).

\bibitem{Lindblad1976}
G. Lindblad.
\newblock ``On the generators of quantum dynamical semigroups''.
\newblock \href{https://doi.org/10.1007/BF01608499}{Comm. Math. Phys. {\bf 48}, 119--130}~(1976).

\bibitem{Brody_2013}
D.~C. Brody.
\newblock ``Biorthogonal quantum mechanics''.
\newblock \href{https://doi.org/10.1088/1751-8113/47/3/035305}{J. Phys. A {\bf 47}, 035305}~(2013).

\bibitem{thermodynamics}
Bart{\l}omiej Gardas, Sebastian Deffner, and Avadh Saxena.
\newblock ``Non-hermitian quantum thermodynamics''.
\newblock \href{https://doi.org/10.1038/srep23408}{Sci. Rep. {\bf 6}, 23408}~(2016).

\bibitem{DaJian2019b}
D.-J. Zhang, Q.-H. Wang, and J. Gong.
\newblock ``Time-dependent $\mathcal{PT}$-symmetric quantum mechanics in generic non-hermitian systems''.
\newblock \href{https://doi.org/10.1103/PhysRevA.100.062121}{Phys. Rev. A {\bf 100}, 062121}~(2019).

\bibitem{Ju2019}
C.-Y. Ju, A. Miranowicz, G.-Y. Chen, and F. Nori.
\newblock ``Non-Hermitian Hamiltonians and no-go theorems in quantum information''.
\newblock \href{https://doi.org/10.1103/PhysRevA.100.062118}{Phys. Rev. A {\bf 100}, 062118}~(2019).

\bibitem{ZWang-PRL2018}
S. Yao and Z. Wang.
\newblock ``Edge states and topological invariants of non-hermitian systems''.
\newblock \href{https://doi.org/10.1103/PhysRevLett.121.086803}{Phys. Rev. Lett. {\bf 121}, 086803}~(2018).

\bibitem{Heiss_2012}
W.~D. Heiss.
\newblock ``The physics of exceptional points''.
\newblock \href{https://doi.org/10.1088/1751-8113/45/44/444016}{J. Phys. A. {\bf 45}, 444016}~(2012).

\bibitem{review2019EP}
Mohammad-Ali Miri and Andrea Alu.
\newblock ``Exceptional points in optics and photonics''.
\newblock \href{https://doi.org/10.1126/science.aar7709}{Science {\bf 363}, eaar7709}~(2019).

\bibitem{doppler2016dynamically}
J.~Doppler, A.~A. Mailybaev, J.~B{\"o}hm, U.~Kuhl, A.~Girschik, F.~Libisch, T.~J. Milburn, P.~Rabl, N.~Moiseyev, and S.~Rotter.
\newblock ``Dynamically encircling an exceptional point for asymmetric mode switching''.
\newblock \href{https://doi.org/10.1038/nature18605}{Nature {\bf 537}, 76--79}~(2016).

\bibitem{Xu2016}
H.~Xu, D.~Mason, Luyao Jiang, and J.~G.~E. Harris.
\newblock ``Topological energy transfer in an optomechanical system with exceptional points''.
\newblock \href{https://doi.org/10.1038/nature18604}{Nature {\bf 537}, 80--83}~(2016).

\bibitem{nature2017EP}
Hossein Hodaei, Absar~U Hassan, Steffen Wittek, Hipolito Garcia-Gracia, Ramy El-Ganainy, Demetrios~N Christodoulides, and Mercedeh Khajavikhan.
\newblock ``Enhanced sensitivity at higher-order exceptional points''.
\newblock \href{https://doi.org/10.1038/nature23280}{Nature {\bf 548}, 187--191}~(2017).

\bibitem{Chen2017}
Weijian Chen, {\c S}ahin Kaya~{\"O}zdemir, Guangming Zhao, Jan Wiersig, and Lan Yang.
\newblock ``Exceptional points enhance sensing in an optical microcavity''.
\newblock \href{https://doi.org/10.1038/nature23281}{Nature {\bf 548}, 192--196}~(2017).

\bibitem{Yoshida}
T. Yoshida, R. Peters, N. Kawakami, and Y. Hatsugai.
\newblock ``Symmetry-protected exceptional rings in two-dimensional correlated systems with chiral symmetry''.
\newblock \href{https://doi.org/10.1103/PhysRevB.99.121101}{Phys. Rev. B {\bf 99}, 121101}~(2019).

\bibitem{Kawabata2019L}
K. Kawabata, T. Bessho, and M. Sato.
\newblock ``Classification of exceptional points and non-{H}ermitian topological semimetals''.
\newblock \href{https://doi.org/10.1103/PhysRevLett.123.066405}{Phys. Rev. Lett. {\bf 123}, 066405}~(2019).

\bibitem{EP:coldatom}
J.~Xu, Y.-X. Du, W.~Huang, and D.-W. Zhang.
\newblock ``Detecting topological exceptional points in a parity-time symmetric system with cold atoms''.
\newblock \href{https://doi.org/10.1364/OE.25.015786}{Opt. Express {\bf 25}, 15786--15795}~(2017).

\bibitem{Peng2016}
Bo~Peng, {\c S}ahin~Kaya {\"O}zdemir, Matthias Liertzer, Weijian Chen, Johannes Kramer, Huzeyfe Y{\i}lmaz, Jan Wiersig, Stefan Rotter, and Lan Yang.
\newblock ``Chiral modes and directional lasing at exceptional points''.
\newblock \href{https://doi.org/10.1073/pnas.1603318113}{Proceedings of the National Academy of Sciences {\bf 113}, 6845--6850}~(2016).

\bibitem{Zhou2018}
H. Zhou, C. Peng, Y. Yoon, C.~W. Hsu, K.~A. Nelson, Liang Fu, J.~D. Joannopoulos, Marin Solja{\v c}i{\'c}, and Bo~Zhen.
\newblock ``Observation of bulk {F}ermi arc and polarization half charge from paired exceptional points''.
\newblock \href{https://doi.org/10.1126/science.aap9859}{Science {\bf 359}, 1009--1012}~(2018).

\bibitem{PRXQuantum_2021}
G.-Q. Zhang, Z. Chen, D.~Xu, N. Shammah, M. Liao, T.-F. Li, L. Tong, S.-Y. Zhu, F. Nori, and J.~Q. You.
\newblock ``Exceptional point and cross-relaxation effect in a hybrid quantum system''.
\newblock \href{https://doi.org/10.1103/PRXQuantum.2.020307}{PRX Quantum {\bf 2}, 020307}~(2021).

\bibitem{magnondevice_Sci2019}
H. Liu, D. Sun, C. Zhang, M. Groesbeck, R. Mclaughlin, and Z~Valy Vardeny.
\newblock ``Observation of exceptional points in magnonic parity-time symmetry devices''.
\newblock \href{https://doi.org/10.1126/sciadv.aax9144}{Science Advances {\bf 5}, eaax9144}~(2019).

\bibitem{Bender1998}
C.~M. Bender and S. Boettcher.
\newblock ``Real spectra in non-{H}ermitian {H}amiltonians having PT symmetry''.
\newblock \href{https://doi.org/10.1103/PhysRevLett.80.5243}{Phys. Rev. Lett. {\bf 80}, 5243--5246}~(1998).

\bibitem{Bender2002}
C.~M. Bender, D.~C. Brody, and H.~F. Jones.
\newblock ``Complex extension of quantum mechanics''.
\newblock \href{https://doi.org/10.1103/PhysRevLett.89.270401}{Phys. Rev. Lett. {\bf 89}, 270401}~(2002).

\bibitem{Bender2004}
C.~M. Bender, J. Brod, Andr{\'{e}} Refig, and Moretz~E Reuter.
\newblock ``The {C} operator in {PT}-symmetric quantum theories''.
\newblock \href{https://doi.org/10.1088/0305-4470/37/43/009}{J. Phys. A: Math. Gen. {\bf 37}, 10139--10165}~(2004).

\bibitem{Bender2007}
C.~M. Bender.
\newblock ``Making sense of non-{H}ermitian {H}amiltonians''.
\newblock \href{https://doi.org/10.1088/0034-4885/70/6/r03}{Rep. Prog. Phys. {\bf 70}, 947--1018}~(2007).

\bibitem{Ashida2017}
Y. Ashida, S. Furukawa, and M. Ueda.
\newblock ``Parity-time-symmetric quantum critical phenomena''.
\newblock \href{https://doi.org/10.1038/ncomms15791}{Nature Communications {\bf 8}, 15791}~(2017).

\bibitem{ozturk2021observation}
Fahri~Emre {\"O}zt{\"u}rk, Tim Lappe, G{\"o}ran Hellmann, Julian Schmitt, Jan Klaers, Frank Vewinger, Johann Kroha, and Martin Weitz.
\newblock ``Observation of a non-{H}ermitian phase transition in an optical quantum gas''.
\newblock \href{https://doi.org/10.1126/science.abe9869}{Science {\bf 372}, 88--91}~(2021).

\bibitem{Gong2018}
Z. Gong, Y. Ashida, K. Kawabata, K. Takasan, S. Higashikawa, and M. Ueda.
\newblock ``Topological phases of non-{H}ermitian systems''.
\newblock \href{https://doi.org/10.1103/PhysRevX.8.031079}{Phys. Rev. X {\bf 8}, 031079}~(2018).

\bibitem{Kawabata2019X}
K. Kawabata, K. Shiozaki, M. Ueda, and M. Sato.
\newblock ``Symmetry and topology in non-{H}ermitian physics''.
\newblock \href{https://doi.org/10.1103/PhysRevX.9.041015}{Phys. Rev. X {\bf 9}, 041015}~(2019).

\bibitem{QPT-XY}
Y.-G. Liu, L.~Xu, and Z. Li.
\newblock ``Quantum phase transition in a non-{H}ermitian {XY} spin chain with global complex transverse field''.
\newblock \href{https://doi.org/10.1088/1361-648x/ac00dd}{J. Phys.: Cond. Mat. {\bf 33}, 295401}~(2021).

\bibitem{Longhi_2019F}
S. Longhi.
\newblock ``Loschmidt echo and fidelity decay near an exceptional point''.
\newblock \href{https://doi.org/10.1002/andp.201900054}{Annalen der Physik {\bf 531}, 1900054}~(2019).

\bibitem{Sun2022EPL}
J.-C. Tang, S.-P. Kou, and G. Sun.
\newblock ``Dynamical scaling of {L}oschmidt echo in non-{H}ermitian systems''.
\newblock \href{https://doi.org/10.1209/0295-5075/ac53c4}{Europhysics Letters {\bf 137}, 40001}~(2022).

\bibitem{Banerjee_2021}
A. Banerjee and A. Narayan.
\newblock ``Non-{H}ermitian semi-{D}irac semi-metals''.
\newblock \href{https://doi.org/10.1088/1361-648x/abe796}{J. Phys.: Cond. Mat. {\bf 33}, 225401}~(2021).

\bibitem{PYChang2020}
P.-Y. Chang, J.-S. You, X. Wen, and S. Ryu.
\newblock ``Entanglement spectrum and entropy in topological non-{H}ermitian systems and nonunitary conformal field theory''.
\newblock \href{https://doi.org/10.1103/PhysRevResearch.2.033069}{Phys. Rev. Research {\bf 2}, 033069}~(2020).

\bibitem{Pires-PRB2021}
D.~P. Pires and T. Macr\`{\i}.
\newblock ``Probing phase transitions in non-{H}ermitian systems with multiple quantum coherences''.
\newblock \href{https://doi.org/10.1103/PhysRevB.104.155141}{Phys. Rev. B {\bf 104}, 155141}~(2021).

\bibitem{tu2021renyi}
Y.-T. Tu, Y.-C. Tzeng, and P.-Y. Chang.
\newblock ``R\'enyi entropies and negative central charges in non-{H}ermitian quantum systems''.
\newblock \href{https://doi.org/10.21468/SciPostPhys.12.6.194}{SciPost Physics {\bf 12}, 194}~(2022).

\bibitem{scar}
Qianqian Chen, Shuai~A. Chen, and Zheng Zhu.
\newblock ``Weak ergodicity breaking in non-{H}ermitian many-body systems''~(2022).
\newblock \href{http://arxiv.org/abs/2202.08638}{arXiv:2202.08638}.

\bibitem{wang2022non}
Y.-C. Wang, J.-S. You, and H.-H. Jen.
\newblock ``A non-{H}ermitian optical atomic mirror''.
\newblock \href{https://doi.org/10.1038/s41467-022-32372-3}{Nature Communications {\bf 13}, 4598}~(2022).

\bibitem{Tzeng2021}
Y.-C. Tzeng, C.-Y. Ju, G.-Y. Chen, and W.-M. Huang.
\newblock ``Hunting for the non-{H}ermitian exceptional points with fidelity susceptibility''.
\newblock \href{https://doi.org/10.1103/PhysRevResearch.3.013015}{Phys. Rev. Research {\bf 3}, 013015}~(2021).

\bibitem{fsus-RR_PRL_2020}
N. Matsumoto, K. Kawabata, Y. Ashida, S. Furukawa, and M. Ueda.
\newblock ``Continuous phase transition without gap closing in non-{H}ermitian quantum many-body systems''.
\newblock \href{https://doi.org/10.1103/PhysRevLett.125.260601}{Phys. Rev. Lett. {\bf 125}, 260601}~(2020).

\bibitem{half}
H. Jiang, C. Yang, and S. Chen.
\newblock ``Topological invariants and phase diagrams for one-dimensional two-band non-{H}ermitian systems without chiral symmetry''.
\newblock \href{https://doi.org/10.1103/PhysRevA.98.052116}{Phys. Rev. A {\bf 98}, 052116}~(2018).

\bibitem{DaJian2019a}
D.-J. Zhang, Q.-H. Wang, and J. Gong.
\newblock ``Quantum geometric tensor in $\mathcal{PT}$-symmetric quantum mechanics''.
\newblock \href{https://doi.org/10.1103/PhysRevA.99.042104}{Phys. Rev. A {\bf 99}, 042104}~(2019).

\bibitem{fsus:perturbation}
G.~Sun, J.-C. Tang, and S.-P. Kou.
\newblock ``Biorthogonal quantum criticality in non-{H}ermitian many-body systems''.
\newblock \href{https://doi.org/10.1007/s11467-021-1126-1}{Frontiers of Physics {\bf 17}, 33502}~(2022).

\bibitem{Chen_2019}
C. Chen, L. Jin, and R.-B. Liu.
\newblock ``Sensitivity of parameter estimation near the exceptional point of a non-{H}ermitian system''.
\newblock \href{https://doi.org/10.1088/1367-2630/ab32ab}{New J. Phys. {\bf 21}, 083002}~(2019).

\bibitem{nishiyama2020imaginary}
Y. Nishiyama.
\newblock ``Imaginary-field-driven phase transition for the 2d ising antiferromagnet: A fidelity-susceptibility approach''.
\newblock \href{https://doi.org/10.1016/j.physa.2020.124731}{Physica A {\bf 555}, 124731}~(2020).

\bibitem{nishiyama2020fidelity}
Y. Nishiyama.
\newblock ``Fidelity-susceptibility analysis of the honeycomb-lattice ising antiferromagnet under the imaginary magnetic field''.
\newblock \href{https://doi.org/10.1140/epjb/e2020-10264-5}{Eur. Phys. J. B {\bf 93}, 1--7}~(2020).

\bibitem{fsus-RR_EPL_2020}
C.-X. Guo, X.-R. Wang, and S.-P. Kou.
\newblock ``Non-{H}ermitian avalanche effect: Non-perturbative effect induced by local non-{H}ermitian perturbation on a {Z}$_2$ topological order''.
\newblock \href{https://doi.org/10.1209/0295-5075/131/27002}{Europhysics Letters {\bf 131}, 27002}~(2020).

\bibitem{Macri:2021}
Ygor Par\'a, Giandomenico Palumbo, and Tommaso Macr\`{\i}.
\newblock ``Probing non-{H}ermitian phase transitions in curved space via quench dynamics''.
\newblock \href{https://doi.org/10.1103/PhysRevB.103.155417}{Phys. Rev. B {\bf 103}, 155417}~(2021).

\bibitem{perturbation1972}
M.~M. Sternheim and J.~F. Walker.
\newblock ``Non-{H}ermitian {H}amiltonians, decaying states, and perturbation theory''.
\newblock \href{https://doi.org/10.1103/PhysRevC.6.114}{Phys. Rev. C {\bf 6}, 114--121}~(1972).

\bibitem{intrinsic}
S. Chen, L.~Wang, Y. Hao, and Y. Wang.
\newblock ``Intrinsic relation between ground-state fidelity and the characterization of a quantum phase transition''.
\newblock \href{https://doi.org/10.1103/PhysRevA.77.032111}{Phys. Rev. A {\bf 77}, 032111}~(2008).

\bibitem{perturbation_JPCM_2021}
A. Marie, Hugh G~A Burton, and Pierre-Fran{\c{c}}ois Loos.
\newblock ``Perturbation theory in the complex plane: exceptional points and where to find them''.
\newblock \href{https://doi.org/10.1088/1361-648x/abe795}{J. Phys.: Cond. Mat. {\bf 33}, 283001}~(2021).

\bibitem{SSH1979}
W.~P. Su, J.~R. Schrieffer, and A.~J. Heeger.
\newblock ``Solitons in polyacetylene''.
\newblock \href{https://doi.org/10.1103/PhysRevLett.42.1698}{Phys. Rev. Lett. {\bf 42}, 1698--1701}~(1979).

\bibitem{Tzeng2016}
Y.-C. Tzeng, L.~Dai, M.~Chung, Luigi Amico, and Leong-Chuan Kwek.
\newblock ``Entanglement convertibility by sweeping through the quantum phases of the alternating bonds {XXZ} chain''.
\newblock \href{https://doi.org/10.1038/srep26453}{Sci. Rep. {\bf 6}, 26453}~(2016).

\bibitem{Tzeng2020}
Y.-C. Tzeng and M.-F. Yang.
\newblock ``Fate of Fermi-arc states in gapped {W}eyl semimetals under long-range interactions''.
\newblock \href{https://doi.org/10.1103/PhysRevB.102.035148}{Phys. Rev. B {\bf 102}, 035148}~(2020).

\bibitem{nSSH_optical_2018}
M.~Parto, S.~Wittek, H.~Hodaei, G.~Harari, M.~A. Bandres, J.~Ren, M.~C. Rechtsman, M.~Segev, D.~N. Christodoulides, and M.~Khajavikhan.
\newblock ``Edge-mode lasing in 1D topological active arrays''.
\newblock \href{https://doi.org/10.1103/PhysRevLett.120.113901}{Phys. Rev. Lett. {\bf 120}, 113901}~(2018).

\bibitem{pan2018photonic}
M.~Pan, H.~Zhao, P.~Miao, S.~Longhi, and L.~Feng.
\newblock ``Photonic zero mode in a non-hermitian photonic lattice''.
\newblock \href{https://doi.org/10.1038/s41467-018-03822-8}{Nat. Comm. {\bf 9}, 1--8}~(2018).

\bibitem{Song_2019}
W. Song, W. Sun, C. Chen, Q. Song, S. Xiao, S. Zhu, and T. Li.
\newblock ``Breakup and recovery of topological zero modes in finite non-{H}ermitian optical lattices''.
\newblock \href{https://doi.org/10.1103/PhysRevLett.123.165701}{Phys. Rev. Lett. {\bf 123}, 165701}~(2019).

\bibitem{nSSH_es1}
L.~Herviou, N.~Regnault, and J.~H. Bardarson.
\newblock ``{Entanglement spectrum and symmetries in non-Hermitian fermionic non-interacting models}''.
\newblock \href{https://doi.org/10.21468/SciPostPhys.7.5.069}{SciPost Physics {\bf 7}, 69}~(2019).

\bibitem{nSSH}
S.~Lieu.
\newblock ``Topological phases in the non-{H}ermitian {S}u-{S}chrieffer-{H}eeger model''.
\newblock \href{https://doi.org/10.1103/PhysRevB.97.045106}{Phys. Rev. B {\bf 97}, 045106}~(2018).

\bibitem{Mikeska2004}
Hans-J{\"u}rgen Mikeska and Alexei~K. Kolezhuk.
\newblock ``One-dimensional magnetism''.
\newblock \href{https://doi.org/10.1007/BFb0119591}{Pages 1--83}.
\newblock Springer Berlin Heidelberg. Berlin, Heidelberg~(2004).

\bibitem{Tzeng2012}
Y.-C. Tzeng.
\newblock ``Parity quantum numbers in the density matrix renormalization group''.
\newblock \href{https://doi.org/10.1103/PhysRevB.86.024403}{Phys. Rev. B {\bf 86}, 024403}~(2012).

\bibitem{TensorNetwork}
H.-Q. Zhou, R. Or\'us, and G. Vidal.
\newblock ``Ground state fidelity from tensor network representations''.
\newblock \href{https://doi.org/10.1103/PhysRevLett.100.080601}{Phys. Rev. Lett. {\bf 100}, 080601}~(2008).

\bibitem{Liang2013}
S.-D. Liang and G.-Y. Huang.
\newblock ``Topological invariance and global berry phase in non-hermitian systems''.
\newblock \href{https://doi.org/10.1103/PhysRevA.87.012118}{Phys. Rev. A {\bf 87}, 012118}~(2013).

\end{thebibliography}

\section*{Appendix}
\appendix

\section{Right-right fidelity susceptibility in SSH ladder} 
The right-right fidelity in the SSH model Eq.~\eqref{eq:sshHamiltonian} is defined by considering the conventional inner-product of the right eigenvectors only, $\mathcal{F}^{RR}=\abs{\langle\psi_0^R(v_1)|\psi_0^R(v_1+\epsilon)\rangle}^2$.
The single-particle fidelity susceptibility for $\Delta_k>0$ is
$\chi_k^{\rm RR}=-\frac{\Omega}{8\Delta_k(2v_1v_2\cos2k+2\cos k(v_1+v_2)+v_1^2+v_2^2+1)^2}$,
where
$\Omega=4 \sqrt{\Delta_k}uv_1v_2\sin2k+4\sqrt{\Delta_k}uv_1\sin k
   +2\sqrt{\Delta_k}uv_2^2\sin4k+4\sqrt{\Delta_k}uv_2\sin3k
    +2\sqrt{\Delta_k}u\sin2k+v_2\cos4k(v_2 (-2 u^2+v_1^2+3)+3 v_1+v_2^3)
    -\cos k(4u^2v_1+2v_1^2v_2+2v_1v_2^2+v_1+4v_2^3+3v_2)
+\cos2k(-2u^2(2v_1v_2+1)+v_1^2-v_1v_2(v_2^2+2)+v_2^2+1
+\cos3k(v_2(-4 u^2+2 v_1^2+3)-v_1 v_2^2+v_1+3 v_2^3)-v_1^2 (2 u^2+v_2^2+1)
+v_1 v_2^3 \cos 6 k+v_2^2 \cos 5 k (3 v_1+v_2)-v_1 v_2-(v_2^2+4) v_2^2-1$,
and for $\Delta_k<0$,
$\chi_k^{\rm RR}=\frac{\Upsilon}{8\Delta_ku^2}$, where $\Upsilon=-[2 u^2+v_2 (v_2 \cos 4 k-2 \cos k+2 \cos 3 k)+\cos 2 k-v_2^2-1]$.

\begin{figure}[t]
\center
\includegraphics[width=4.2in]{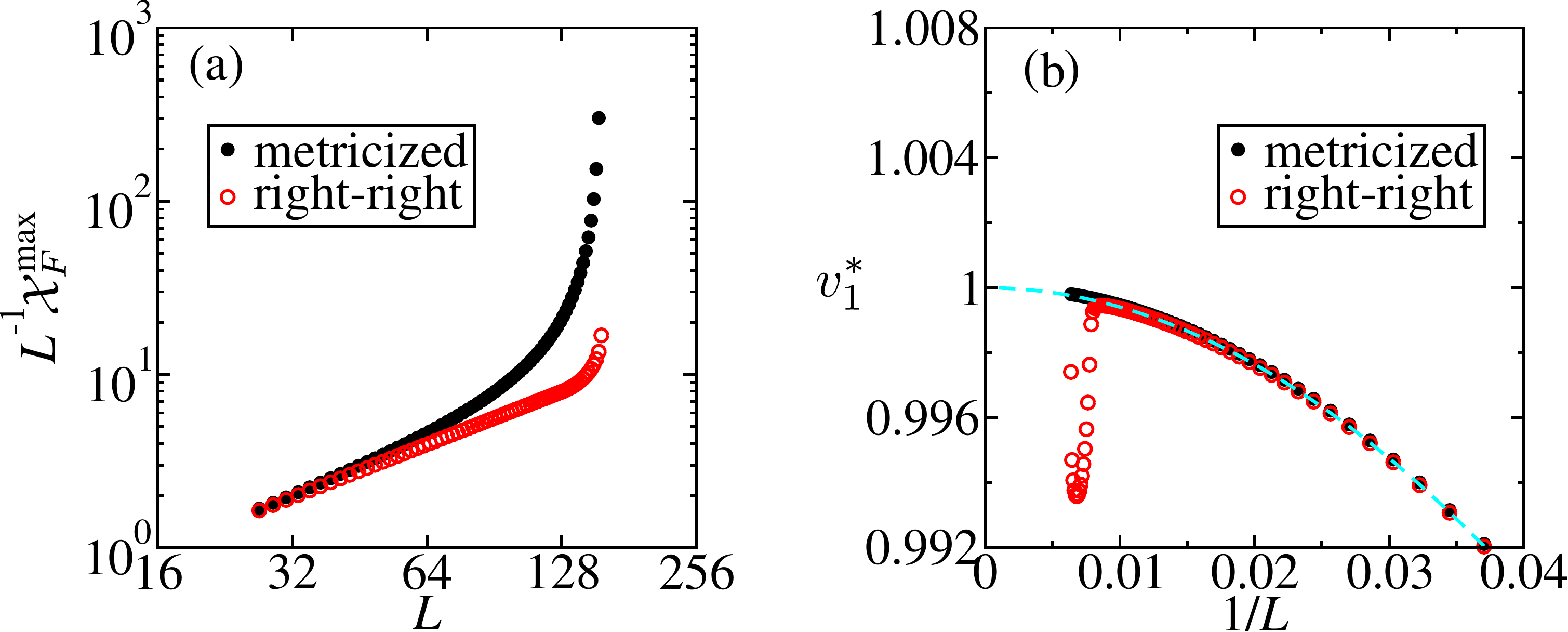}
\caption{Finite size scaling of the fidelity susceptibility density of the SSH model for fixed $v_2=0$ and $u=0.02$, i.e. the first red regime in Fig.~\ref{fig:SSHDensityPlot}(left). The system sizes are chosen to be small enough $L<L_0$ such that the ground state is still PT-unbroken with real total energy. (a) The log-log plot of the peak value shows a faster-than-power-law divergence with the system size. (b) The extrapolation of the peak position shows that the value is close to the Hermitian quantum critical point $v_1=1$.}\label{fig:odd-scaling-ssh}
\end{figure}

\section{Enhancement of fidelity susceptibility}
From the perspective of the energy spectrum, both the quantum critical point and the EP are the gap closing points. The gap closing takes place in the thermodynamic limit for the Hermitian quantum phase transition. The finite size system remains a finite size gap even at the quantum critical point. 
However, the EP does not require the thermodynamic limit. Therefore, when the non-Hermitian term tries to drive the finite size system at the critical point into the PT-broken phase, the gap closing happens early in the finite size due to the parameter going through the EP. The signal of the phase transition is thus amplified by the small non-Hermitian term.
The enhancement of the fidelity susceptibility for the quantum phase transition by a small non-Hermitian term in the finite size system is found, as the first red regime shown in Fig.~\ref{fig:SSHDensityPlot} for both $v_2=0$ and $v_2=\frac{1}{2}(1+\sqrt{5})$.
Furthermore, the enhanced positive peaks extend into the PT-broken phase, as the red dashed curves shown in Fig.~\ref{fig:SSHDensityPlot}.
It is suggested that this enhancement is helpful for detecting phase transitions in finite size systems. Here, we perform the finite size scaling in the first red regime.

In Fig.~\ref{fig:odd-scaling-ssh}(a), we show that for the fixed $u=0.02$ and $v_2=0$, both the metricized and the right-right fidelity susceptibility density are enhanced, with a faster-than-power-law divergence with the system size. In Fig.~\ref{fig:odd-scaling-ssh}(b), the extrapolation of the peak position shows that the value from the metricized fidelity susceptibility density is close to the Hermitian quantum critical point $v_1=1$.

Although the enhancement of the peak by the non Hermitian term may not be a general phenomenon, it is expected the similar behavior can happen in the general non-interacting two band models.
For general non-interacting two band models, it is necessary to close a real energy gap to have a pair of PT-broken states due to PT-symmetry. Therefore the gap becomes closer as a non-Hermitian term is introduced. For example, in our SSH model, we introduce a non-Hermitian term and it makes two bands of all momentum points be closer which can be seen clearly from the energy value $\varepsilon_{\pm}(k)=\pm\sqrt{\Delta_k}$. Even there is a momentum dependent non-Hermitian term, e.g. a non-Hermitian hopping term, this term still induces two bands to be closer. Since the effect of this non-Hermitian term is momentum dependent, as a result, it may induce a gap closing at a momentum point which is different from the original gap-closing momentum point in the quantum phase transition. 
However, the momentum shift is expected to be small as the non-Hermitian term is a small perturbation. 
As a result, a small non-Hermitian term may still enhance the signal of Hermitian quantum phase transition, but in general the finite size scaling of the enhancement of fidelity susceptibility becomes complicated.

\section{Boundary modes and complex Berry phase in the PT symmetric SSH model}
\label{app:Boundary}
\begin{figure}[t]
\center
    \includegraphics[width=0.9\textwidth]{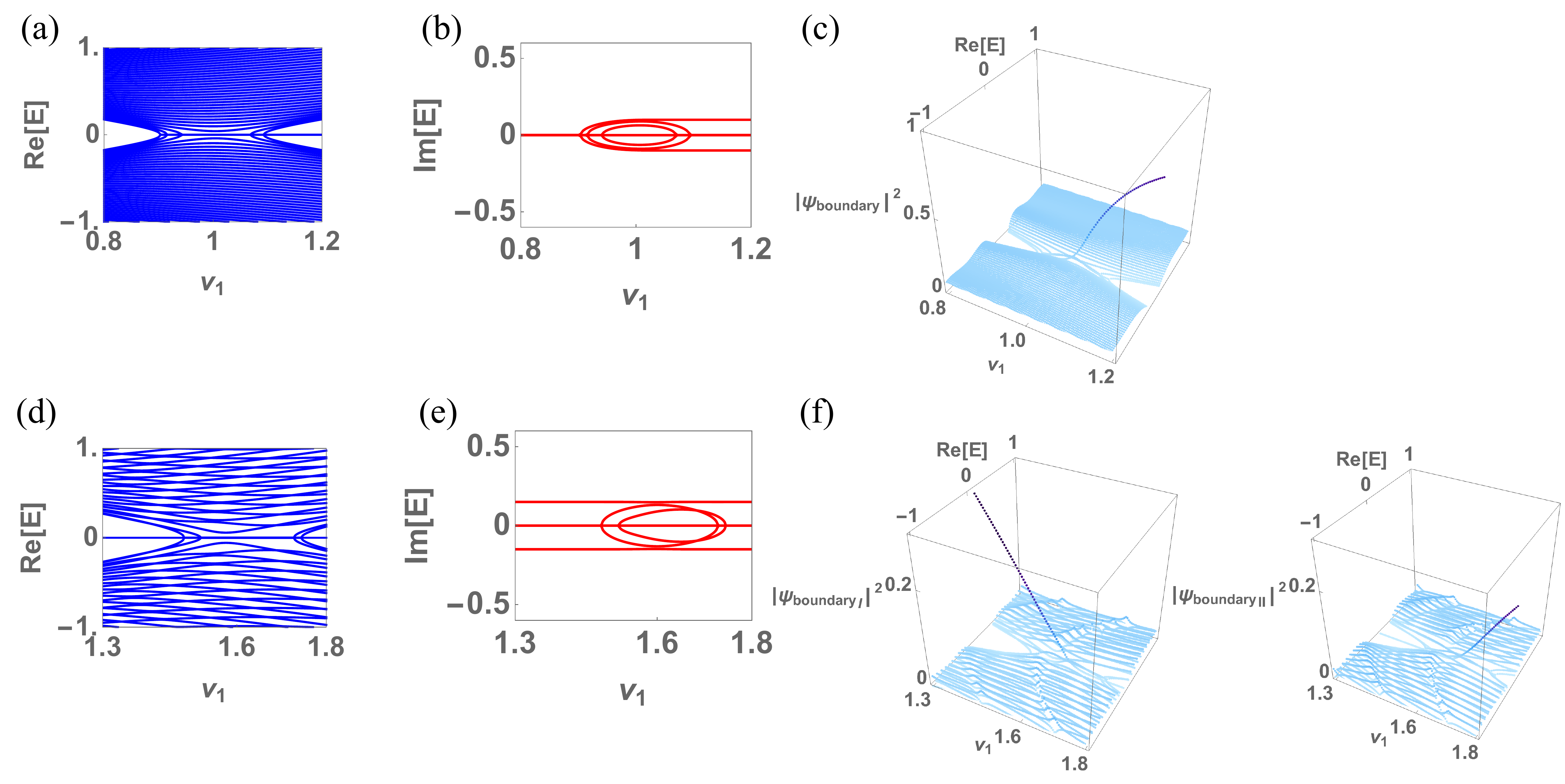}
    \caption{$v_2=0$, $u=0.1$: (a) Real part of the energy, (b) imaginary part of the energy and (c) 3D plot of the  boundary state amplitude.
    $v_2=\frac{1+\sqrt{5}}{2}$, $u=0.15$: (d) Real part of the energy, (e) imaginary part of the energy and (f) 3D plot of the  boundary state amplitude [left/right panel corresponds to boundary mode $I/II$].}
    \label{fig:Top_boundary}
\end{figure}

\begin{figure}[t]
\center
    \includegraphics[width=0.6\textwidth]{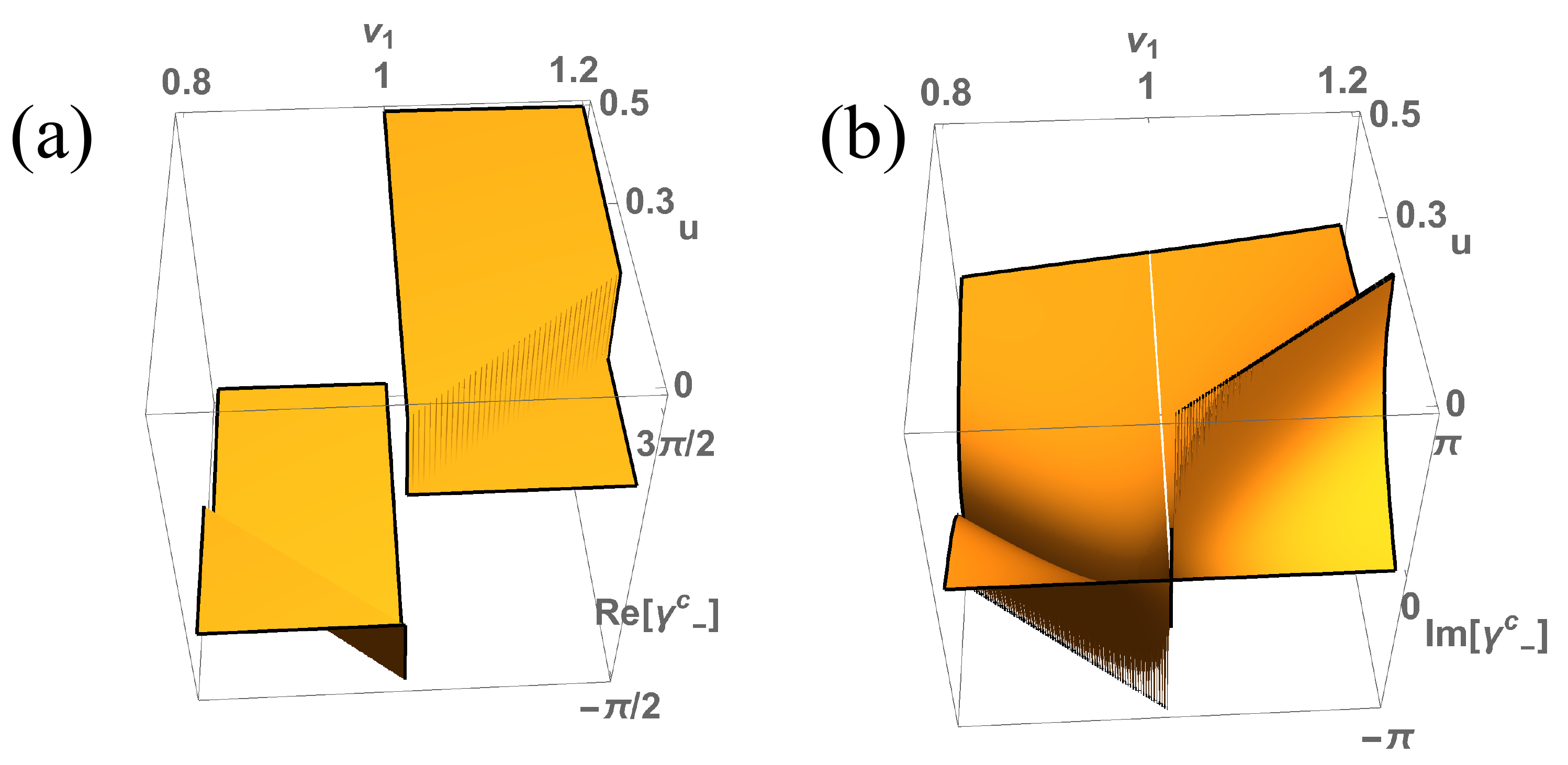}
    \caption{(a) Real part and (b) complex part of the complex Berry phase for the occupied state $\gamma^c_{-} $ as a function of $v_1$ and $u$ with $v_2=0$ and $w=1$. There is a $2 \pi$ jump in the real part of $\gamma^c_{-} $ at $v_1=1$ which is the phase boundary between
     the topological and trivial phases. There are divergences in the imaginary part of $\gamma^c_{-} $ which separate the PT-symmetric and PT-broken phases. }
    \label{fig:cBP}
\end{figure}
We compute the spectrum of the PT symmetric SSH model with open boundary condition and identify the topological property from the existence of the boundary modes.
As shown in Figs.~\ref{fig:Top_boundary}(c) and (f), the boundary modes can be clearly separated from the bulk modes in the PT broken region.
We find the phase boundary between the topological and trivial phases is independent of the imaginary chemical potential $u$ as it discussed in Ref.~\cite{nSSH}.
In Fig.~\ref{fig:Top_boundary}(f), there are two distinct topological phases where the one phase has a localized boundary mode at sublattice $\uparrow$ site  and the other has a localized boundary mode at
sublattice $\downarrow$ site.

We also compute the complex Berry phase~\cite{nSSH, Liang2013},
\begin{align}
\gamma^c_{pm} = \oint_k A_{\pm},
\end{align}
where $A_{\pm} = i \langle L_{\pm}(k) |d_k R_{\pm} (k) \rangle$ with $|L/R_\pm(k) \rangle$ defined in Eqs.~(\ref{eq:SSHreEeigen}) and (\ref{eq:SSHimEeigen}).
Following from Ref.~\cite{Liang2013}, we can obtain an analytic form of the complex Berry phase for $v_2=0$
\begin{align}
\gamma^c_{\pm}  = \pi \Theta (v_1/w -1) \mp i \frac{u}{2 w} \sqrt{\frac{y w}{v_1}}\left( K(y) + \frac{v_1 - w}{v_1 +w} \Pi (x,y) \right),
\end{align}
where $K(y) = \int_0^{\pi/2} dz \frac{1}{\sqrt{1- y \sin^2 z}}$  and $\Pi(x,y) =  \int_0^{\pi/2} dz \frac{1}{(1-x \sin^2 z)\sqrt{1- y \sin^2 z}} $
are the first and the third kinds of the complete elliptic integrals with $x = \frac{4 v_1}{w (v_1/w +1)^2}$ and $y = \frac{4 v_1/ w}{(v_1/w +1)^2 - u^2/w^2}$.
In Fig.~\ref{fig:cBP}, the real part of $\gamma^c_{-}$ has a $2\pi $ jump at $v_1 =1$ which is the phase boundary of the topological and trivial phases. This feature agrees with 
the existence boundary modes in the energy spectrum with open boundary condition. There are also kinks in the real part of $\gamma^c_{-}$ which are the phase boundaries between PT-symmetric and PT-broken phase.
The imaginary part of $\gamma^c_{-}$ diverges at these boundaries. 

\end{document}